\documentclass[preprint,12pt]{elsarticle}

\usepackage{epsf,epsfig,float,amssymb,latexsym}
\usepackage{rotating}
\usepackage{xcolor,url,ulem}
\usepackage{graphicx}
\usepackage{lscape}
\usepackage{amsmath,dsfont}
\usepackage{amssymb}

\newlength{\MySep}
\newcommand {\beq}     {\begin{equation}}
\newcommand {\eeq}[1]  {\label{#1}\end{equation}}
\newcommand {\beqa}    {\begin{eqnarray}}
\newcommand {\eeqa}[1] {\label{#1}\end{eqnarray}}
\newcommand {\eeqan}   {\end{eqnarray}}
\newcommand {\CR}      {\nonumber \\ }

\newcommand {\BeNul}       {b_{0}}
\newcommand {\BeDe}        {b_{\rm D}}
\newcommand {\BeFe}        {b_{\rm F}}

\newcommand {\DxD}         {D^{2}}
\newcommand {\DxF}         {DF}
\newcommand {\FxF}         {F^{2}}

\makeatletter

\def\compoundrel#1\over#2{\mathpalette\compoundreL{{#1}\over{#2}}}
\def\compoundreL#1#2{\compoundREL#1#2}
\def\compoundREL#1#2\over#3{\mathrel
      {\vcenter{\hbox{$\m@th\buildrel{#1#2}\over{#1#3}$}}}}
\makeatother

\journal{Nuclear Physics A}

\begin{document}

\begin{frontmatter}



\title{Coupled channels approach to $\eta N$ and $\eta' N$ interactions}

\author[UJF]{P.~C.~Bruns}
\author[UJF]{A.~Ciepl\'{y} \corref{correspondence}}
\cortext[correspondence]{Corresponding author}
\ead{cieply@ujf.cas.cz}

\address[UJF]{Nuclear Physics Institute of the Czech Academy of Sciences, 250 68 \v{R}e\v{z}, Czech Republic}

\begin{abstract}
We present a coupled channels separable potential approach to $\eta N$ and $\eta'N$ 
interactions using a chiral-symmetric interaction kernel. The s-wave $\pi N$ amplitudes 
and $\pi^{-}p$ induced total cross sections are reproduced satisfactorily in a broad 
interval of energies despite limiting the channel space to two-body interactions 
of pseudoscalar mesons with the baryon ground-state octet. It is demonstrated that 
an explicit inclusion of the $\eta_0$ meson singlet field leads to a more attractive 
$\eta N$ interaction, with the real part of the scattering length exceeding 1 fm. 
The $\eta'N$ diagonal coupling appears sufficient to generate an $\eta'N$ bound state 
but the inter-channel dynamics moves the respective pole far from physical region 
making the $\eta'N$ interaction repulsive at energies around the channel threshold. 
The $N^{*}(1535)$ and $N^{*}(1650)$ resonances are generated dynamically 
and the origin and properties of the $S$-matrix poles assigned to them are studied 
in detail. We also hint at a chance that the $N^{*}(1895)$ state might 
also be formed provided a suitably varied model setting is found.
\end{abstract}

\begin{keyword}
chiral dynamics \sep meson-nucleon interaction \sep eta-eta' mixing \sep baryon resonances 
\end{keyword}

\end{frontmatter}


\section{Introduction}
\label{sec:intro}

Modern treatments of meson--baryon interactions at low energies are based on chiral perturbation 
theory (ChPT) that implements the QCD symmetries in its nonperturbative regime. Based on the method 
of ``phenomenological Lagrangians'' \cite{Weinberg:1978kz}, corrections to the predictions 
of current algebra can be systematically computed order by order in an expansion of QCD Green functions 
in powers of small momenta and light quark masses \cite{Gasser:1984gg}, respecting the chiral symmetry of QCD
as well as other field-theoretic constraints. Naturally, the effective theory is expected to work well in the SU(2) 
sector due to smallness of the {\it up} and {\it down} quark masses. However, when including 
the baryon ground-state octet in the effective Lagrangian \cite{Gasser:1987rb,Krause:1990xc} 
one often faces the problem of a bad convergence behaviour of the low-energy expansion. 
The situation calls for non-perturbative extensions of the standard effective-field-theory framework, 
usually entailing resummations of certain higher-order corrections. The unavoidable model-dependence
of such approaches can be controlled, to some extent, by implementing constraints from chiral symmetry.
Interestingly, an application of this method to the three-flavor sector of meson-baryon scattering 
has lead to a very successful description of $\bar{K}N$ interactions despite the relatively large mass 
of the strange quark and the presence of the $\Lambda(1405)$ resonance just below the $\bar{K}N$ threshold  
see e.g.~the pioneering works \cite{Kaiser:1995eg, Oset:1997it}. There, ChPT is supplemented 
by a classical resummation technique, the Lippmann-Schwinger equation, which allows to sum up 
the most relevant part of the perturbation series (the rescattering graphs, or ``unitarity corrections''). 
As a result those higher-order corrections 
are accounted for in a situation when the standard perturbation approach does not converge. 
At present, there are several theoretical, chirally motivated approaches on the market that describe 
the multi-channel interactions of the pseudoscalar meson octet ($\pi$, $K$, $\eta$) with the ground
state baryon octet ($N$, $\Lambda$, $\Sigma$, $\Xi$), see e.g.~\cite{Cieply:2016jby} for the recent 
comparison of chiral approaches to the $\bar{K}N$ system and \cite{Inoue:2001ip, Cieply:2013sya, Mai:2012wy} 
for the papers that deal with the $\eta N$ system. 

In the current work, we extend an existing chirally-motivated coupled channels approach for meson-baryon 
scattering in the zero-strangeness sector \cite{Cieply:2013sya} to include an explicit $\eta'$ 
degree of freedom. This is done as a first step to test the applicability of the mentioned model
to a vastly prolonged region of energies, as well as to study the impact of the presence of the $\eta'$ 
on the results reported in \cite{Cieply:2013sya}, and to obtain tentative predictions 
for the $\eta'N \rightarrow \eta'N$ scattering process. As we will demonstrate, the admixture 
of the $\eta_0$ singlet state in the $\eta$ meson makes the $\eta N$ interaction more attractive 
at energies around the channel threshold. This feature is quite relevant for a formation of $\eta$-nuclear 
quasi-bound states as discussed in e.g.~\cite{Inoue:2002xw, Cieply:2013sga, Barnea:2017epo} 
or in the review \cite{Machner:2014ona}.

The treatment of the $\eta'$ as an explicit degree of freedom in the (mesonic) chiral Lagrangian 
has been considered already in \cite{Gasser:1984gg}, see Sec.~12 there, and also in \cite{DiVecchia:1980yfw, 
Kawarabayashi:1980dp, Leutwyler:1996sa, HerreraSiklody:1996pm, Kaiser:2000gs, Beisert:2001qb, Bickert:2016fgy}. 
One reason of interest in the $\eta'$ is that it is prevented from being a ninth Goldstone 
boson (in addition to the pions, kaons and the $\eta_8$) by the axial $U(1)$ anomaly of QCD, 
which however vanishes in the limit of the number of colours $N_{c}$ going to infinity 
\cite{tHooft:1974pnl, Weinberg:1975ui, Witten:1979vv, Coleman:1980mx}. Another reason is 
that the $\eta,\eta'$ mesons can be described as admixtures of a flavor-octet $\eta_8$ and 
a flavor-singlet $\eta_0$ fields, which has some impact e.g.~on the phenomenology of $\eta,\eta'$ 
decays \cite{Fritzsch:1976qc, Gilman:1987ax, Schechter:1992iz, Bramon:1997va, Feldmann:1998vh, 
Feldmann:1999uf, Borasoy:2003yb, Borasoy:2005du,Bijnens:2005sj}. The chiral Lagrangian for baryon ChPT 
has been extended to include an explicit $\eta'$ field in \cite{Bass:1999is, Borasoy:1999nd}. 
In our construction of the meson-baryon potential, we shall follow closely the formulation 
of \cite{Borasoy:2002mt}. In particular, we do not rely on an expansion in $1/N_{c}$ 
(where the mass of the $\eta'$ is to be counted as a small quantity compared to a typical 
hadronic scale $\sim \mathrm{GeV}$), and we will limit ourselves to a simple one-mixing-angle 
scheme for the $\eta,\eta'$ sector, which should be sufficient for a qualitative description
of the observables which we are interested in. 
Moreover, we ignore an additional complication caused by the mixing of the $\pi^{0}$ with 
the $\eta,\eta'$ sector, which represents an isospin-violating effect.

The interaction of the $\eta'$ with baryons is interesting in its own right. Notably, the $\eta'N$ 
scattering length in a free space is an important parameter in assessing the possibility 
of $\eta'$-nucleonic bound states \cite{Tsushima:1998qp, Bass:2005hn, Nagahiro:2011fi, 
Czerwinski:2014yot, Nanova:2016cyn, Tanaka:2016bcp, Metag:2017yuh}. A reduction of the $\eta'$ mass  
in nuclear matter represents another interesting feature related to a partial restoration of chiral symmetry 
and to an in-medium suppression of the $U(1)$ anomaly effects \cite{Pisarski:1983ms, Cohen:1996ng, 
Lee:1996zy, Jido:2011pq}. We refer to \cite{Bass:2018xmz} for a recent review of the $\eta,\eta'$ physics. 
For related studies of $\eta'N$ scattering, see also \cite{Oset:2010ub, Sakai:2014zoa, Anisovich:2018yoo}. 

The paper is organized as follows: In the next section we present the chiral Lagrangian 
and our approach to generating the chirally motivated meson-baryon amplitudes. 
In Section \ref{sec:fits} we discuss the selection of the experimental data and introduce 
several models obtained under various scenarios adopted when fitting the model parameters 
to the data. The main part of the paper, Section \ref{sec:results}, provides our results 
for the fitted observables, the model predictions for the $\eta N$ and $\eta'N$ elastic 
amplitudes, and a discussion of the $S$-matrix poles assigned to the dynamically generated 
$N^{*}(J^{P}=1/2^{-})$ resonant states. The article is closed with a brief summary while 
some lengthy technical points are left for appendices.

\section{Coupled channels chiral model}
\label{sec:model}

\subsection{Chiral U(3) Lagrangians}
\label{sec:Lagr}

We will use the leading order Lagrangians as given in Eqs.~(2-3) of \cite{Borasoy:2002mt}:

\begin{equation}\label{eq:M_LagrBMW}
\mathcal{L}_{M} = \frac{F_{0}^2}{4}\langle u_{\mu}u^{\mu}\rangle + 
\frac{F_{0}^2}{4}\langle\chi_{+}\rangle - 
\frac{v_{0}}{F_{0}^2}\eta_{0}^2 + 
i\frac{v_{3}}{F_{0}}\eta_{0}\langle\chi_{-}\rangle\,,
\end{equation}
\begin{eqnarray}\label{eq:MB_LagrBMW}
  \mathcal{L}^{(1)}_{MB} 
   &=& i\langle\bar{B}\gamma_{\mu}\lbrack D^{\mu},B\rbrack\rangle 
       - \overset{\circ}{m}\langle\bar{B}B\rangle 
       + i\frac{w_{s}}{F_{0}^2}\eta_{0}^2\left(\langle\lbrack D^{\mu},\bar{B}\rbrack\gamma_{\mu}B\rangle 
       - \langle\bar{B}\gamma_{\mu}\lbrack D^{\mu},B\rbrack\rangle\right) \nonumber \\
   &+& \frac{1}{2}D\langle\bar{B}\gamma_{\mu}\gamma_{5}\lbrace u^{\mu},B\rbrace\rangle 
       + \frac{1}{2}F\langle\bar{B}\gamma_{\mu}\gamma_{5}\lbrack u^{\mu},B\rbrack\rangle 
       + \frac{1}{2}D_{s}\langle\bar{B}\gamma_{\mu}\gamma_{5}B\rangle\langle u^{\mu}\rangle\,.
\end{eqnarray}
The ChPT nomenclature and notation used in Eqs.~(\ref{eq:M_LagrBMW}-\ref{eq:MB_LagrBMW}) 
are reviewed in \ref{app:chptnom}. Here we just highlight the two extra terms proportional 
to $w_{s}$ (which was named $u_{1}$ in \cite{Borasoy:2002mt}) and to $D_{s}$, 
that are added to the standard form of the first-order Lagrangian, which 
describes only the coupling of the octet of Goldstone bosons to the 
ground-state baryons \cite{Krause:1990xc}. These two new terms arise due 
to the inclusion of an explicit singlet meson field $\eta_{0}$ added to 
the meson octet to form a meson nonet (compare Eq.~(\ref{eq:Phi}) in \ref{app:chptnom}) 
and leading to $\mathrm{Tr}\,u^{\mu}\equiv\langle 
u^{\mu}\rangle\not=0\,$. In particular, we have an additional axial 
coupling constant $D_{s}$ appearing whenever $\eta_{0}$ couples to a 
baryon, and a baryon-singlet meson contact term $\sim w_{s}$, which is 
not suppressed at low energies by $SU(3)$ chiral symmetry.

As already mentioned in the Introduction, we follow Ref.~\cite{Borasoy:2002mt} and use 
a one-mixing-angle scheme to describe the singlet-octet mixing,
\begin{equation}\label{eq:etamix}
\eta_{8}=\eta\cos\vartheta + \eta'\sin\vartheta\,,\quad \eta_{0}=\eta'\cos\vartheta - \eta\sin\vartheta\,.
\end{equation}
From the leading mesonic Lagrangian of Eq.~(\ref{eq:M_LagrBMW}) one obtains the estimate 
$|\vartheta|\sim 10^{\circ}$ \cite{Gasser:1984gg}. The sign can be determined from the analysis 
of $\eta,\eta'$ decays and comes out negative, see e.g.~\cite{Bramon:1997va}
where a value of $\vartheta\approx -15.5^{\circ}\pm 1.3^{\circ}$ is advocated. 
Note that there are also predictions from lattice QCD, which are now mostly consistent with 
this value, e.g. the result reported in \cite{Christ:2010dd} is $\vartheta = -14.1^{\circ}\pm 2.8^{\circ}$.

At the second chiral order, the relevant terms in the effective meson-baryon Lagrangian 
read (in the notation of Ref.~\cite{Borasoy:2002mt})
\begin{eqnarray}\label{eq:LMB2}
  \mathcal{L}_{MB}^{(2)} 
   &=& b_{D}\langle\bar{B}\lbrace\chi_{+},\,B\rbrace\rangle 
    +  b_{F}\langle\bar{B}\lbrack\chi_{+},\,B\rbrack\rangle 
    +  b_{0}\langle\bar{B}B\rangle\langle\chi_{+}\rangle 
    +  i\frac{c_{D}}{F_{0}}\eta_{0}\langle\bar{B}\lbrace\chi_{-},\,B\rbrace\rangle \nonumber \\
   &+& i\frac{c_{F}}{F_{0}}\eta_{0}\langle\bar{B}\lbrack\chi_{-},\,B\rbrack\rangle 
    +  i\frac{c_{0}}{F_{0}}\eta_{0}\langle\bar{B}B\rangle\langle\chi_{-}\rangle 
    +  d_{1}\langle\bar{B}\lbrace u_{\mu},\,\lbrack u^{\mu},\,B\rbrack\rbrace\rangle \nonumber \\
   &+& d_{2}\langle\bar{B}\lbrack u_{\mu},\,\lbrack u^{\mu},\,B\rbrack\rbrack\rangle 
    +  d_{3}\langle\bar{B}u_{\mu}\rangle\langle u^{\mu}B\rangle 
    +  d_{4}\langle\bar{B}B\rangle\langle u_{\mu}u^{\mu}\rangle \nonumber \\
   &+& d_{5}\langle\bar{B}\lbrace u_{\mu},\,B\rbrace\rangle\langle u^{\mu}\rangle 
    +  d_{6}\langle\bar{B}\lbrack u_{\mu},\,B\rbrack\rangle\langle u^{\mu}\rangle 
    +  d_{7}\langle\bar{B}B\rangle\langle u_{\mu}\rangle\langle u^{\mu}\rangle\,.
\end{eqnarray}
The low energy constants (LECs) $b_{0,D,F}$ match those used in \cite{Cieply:2013sya} 
and in analyses of baryon mass spectra \cite{Jenkins:1991ts,Bernard:1993nj} as well. 
On the Lagrangian (tree graph) level, the LECs $d_{1-4}$ introduced in Eq.~(\ref{eq:LMB2}) 
can be related to the corresponding couplings $d_{D}^{cs},d_{F}^{cs},d_{0}^{cs},d_{1}^{cs},d_{2}^{cs}$ 
used in \cite{Cieply:2013sya} as follows (when setting $d_{2}^{cs}=0$, as the authors 
of the mentioned reference do):
\begin{eqnarray*}
  d_{1} &=& \frac{1}{2}d_{F}^{\,cs}\,,\quad 
  d_{2}  =  \frac{1}{6}d_{D}^{\,cs}\,,\quad 
  d_{3}  =  \frac{1}{6}(3d_{1}^{\,cs}+2d_{D}^{\,cs})\,,\quad 
  d_{4}  =  \frac{1}{6}(3d_{0}^{\,cs}+d_{D}^{\,cs}) \; .
\end{eqnarray*}
Finally, the parameters $c_{0,D,F}$ and $d_{5-7}$ enter through the explicit inclusion 
of the flavor-singlet field $\eta_{0}$. Unfortunately, it turns out that there is not 
enough (sufficiently precise) data in the $\eta N$/$\eta'N$ sector to determine these additional 
subleading LECs reliably. This is not necessarily a big drawback as one can argue that 
(a) the symmetry-breaking LECs $c_{0,D,F}$ are suppressed with respect to $b_{0,D,F}$ 
by a power of $1/N_{c}$, and (b) the $d_{5-7}$ couplings turned out relatively small 
in the fits performed in \cite{Borasoy:2002mt} which included data on meson photoproduction,  
while the fits were stable with respect to small variations of $c_{0,D,F}$ around zero. 
As we will also demonstrate in the current work one can obtain quite satisfactory description 
of the considered data even under the constraint $d_{5-7}=c_{0,D,F}=0$.

\subsection{Meson-baryon scattering amplitudes}
\label{sec:Mod}

In what follows we concentrate on the s-wave meson-baryon interactions and calculate amplitudes 
$f_{0+}^{I}(s)$ in the isospin $I=1/2$ and $I=3/2$ sectors. The notation and 
conventions of \cite{Hoehler:1983lb} are employed here and the coupled channels with zero strangeness are 
ordered according to their thresholds as $|\pi N\rangle$, $|\eta N\rangle$, $|K\Lambda\rangle$, 
$|K\Sigma\rangle$, $|\eta'N\rangle$, see also \ref{app:chptnom} for our phase convention 
concerning the isospin states. The tree-level contributions derived from the effective Lagrangians 
above will yield the potentials $v^{1/2}$ and $v^{3/2}$ of our unitarized scattering 
amplitudes. They are of the form
\begin{equation}\label{eq:f0pPOT1}
f^{I}_{0+,\mathrm{tree}}(s) 
= \frac{\sqrt{E+m}}{F_{\Phi}}\left(\frac{C^{I}(s)}{8\pi\sqrt{s}}\right)\frac{\sqrt{E+m}}{F_{\Phi}} 
=: v^{I}_{0+}(s)\,,
\end{equation}
where we employ a convenient channel-matrix notation in the $5\times 5$ and $2\times 2$ dimensional 
channel spaces for $I=1/2$ and $I=3/2$, respectively. The diagonal matrices $E$, $m$ 
and $F_{\Phi}$ are assembled from the baryon center-of-mass (c.m.) energies, baryon masses and meson decay 
constants of the respective channels, see \ref{app:chanmats}. The channel matrices $C^{I}(s)$ 
contain all the details specific to the effective vertices and the various elastic and inelastic meson-baryon 
reactions. In some more detail,
\begin{eqnarray}
  C(s) \!&=&\! \frac{1}{4}\lbrace(\sqrt{s}-m),\,C_{WT}\rbrace - 2w_{s}\lbrace m,\,C_{w_{s}}\rbrace - C_{ct}^{(2)}(s) \nonumber \\
       \!&-&\! \frac{(\sqrt{s}-m)C_{s}(\sqrt{s}-m)}{\sqrt{s}+m_{N}} - C_{u}(s)  \label{eq:f0pPOT2} \; , \\
  C_{ct}^{(2)}(s) &=& 2M_{\pi}^2 \,C_{\pi} + 2M_{K}^2 \,C_{K} - 2q^{0}(s)\:C_{d}\:q^{0}(s) \; , \nonumber 
\end{eqnarray}
where we omit the isospin superscipts for brevity. The $q^{0}(s)$ represents the diagonal channel matrix 
$q^{0}(s)=(s-m^{2}+M^{2})/(2\sqrt{s})\,$, featuring the meson c.m.~energies in the respective meson-baryon 
channels, while $s$ is the usual Mandelstam variable given by the square of the two-body c.m.~energy. 
The channel matrices $C_{WT}$, $C_{w_{s}}$, $C_{ct}^{(2)}(s)$, $C_{s}$ and $C_{u}(s)$ contain the couplings 
derived from the Weinberg-Tomozawa term of Eq.~(\ref{eq:MB_LagrBMW}), the singlet-term $\sim w_{s}$, 
the contact terms from Eq.~(\ref{eq:LMB2}), and the $s$- and $u$-channel Born terms, respectively. 
Explicit expressions for all coupling matrices can be found in \ref{app:chanmats}. 
In writing Eqs.~(\ref{eq:f0pPOT1}) and (\ref{eq:f0pPOT2}), we have dropped some terms containing 
p-wave projections of invariant amplitudes, which come with factors $\sim E-m$ 
and are suppressed at low energies. The inclusion of the $u$-channel Born graphs in the potential 
requires some subtle modifications in order to avoid violations of unitarity and analyticity. 

The construction of the ``chirally-motivated'' unitarized coupled channels scattering amplitude 
is the same as in \cite{Cieply:2013sya}, and therefore we will be quite brief here. In the model, 
the loop integrals are regulated by the Yamaguchi form factors $g(p)$, featuring regulator scales 
(``soft cutoffs'') $\alpha$ that can also be interpreted as inverse ranges of the interactions. 
For a meson-baryon channel with a baryon $b$ and meson $j$, 
\beq
g_{jb}(p) = \left( 1 + p^{2}/\alpha^{2}_{jb} \right)^{-1}\;.
\eeq{}
The loop functions in this regularization scheme are given by
\beq
G_{jb}(s) = -4\pi\int\frac{d^3p}{(2\pi)^3}\frac{g_{jb}^2(p)}{q_{jb}^2 -p^2 +{\rm i}0} 
          = \frac{(\alpha_{jb}+{\rm i}q_{jb})^2}{2\alpha_{jb}} \,g_{jb}^2(q_{jb}) \;.  
\eeq{eq:Gloop}
Here $q_{jb}\equiv q_{jb}(s)$ is the c.m.~momentum for the meson-baryon channel $jb$, 
see Eq.~(\ref{eq:qcms}). The form factors and loop functions can again be put together 
to form diagonal matrices $g(s)$ and $G(s)$ in the channel space, so that e.g.~$g(s)$ has diagonal 
entries $g_{jb}(q_{jb}(s))$. These matrices are combined with the coupled channels matrix $v_{0+}^{I}$ 
of Eq.~(\ref{eq:f0pPOT1}) to yield our desired model amplitude for isospin $I=1/2$ and $I=3/2$ s-waves:
\beq
f_{0+}^{I}(s) = g(s) \,\lbrack1-v_{0+}^{I}(s)\,G(s)\rbrack^{-1} \,v_{0+}^{I}(s) \,g(s)\,.
\eeq{eq:f0pMOD}
The condition of two-body partial-wave unitarity (in the space of our considered channels) 
can be formulated in a matrix form as follows,
\begin{equation}
\mathrm{Im}\,f_{0+}(s) = f_{0+}^{\ast}(s)\,q(s)\,f_{0+}(s)\,,
\end{equation}
where the $f_{0+}$ matrix comprises the transition amplitudes $[f_{0+}(s)]_{jb,ia}$ and 
$q(s)$ is a diagonal channel matrix assembled from the two-body phase-space factors, 
i.e.~it has entries $[q(s)]_{jb,ia}=\delta_{ab}\,\delta_{ij}\,q_{jb}(s)\,\theta(s-(m_{b}+M_{j})^2)$. 
It is straightforward to verify that the amplitude of Eq.~(\ref{eq:f0pMOD}) fulfills 
the unitarity requirement. 

At this point we find it appropriate to add a remark concerning the relation of our approach 
to the amplitudes and LECs employed in ChPT. Even though our amplitude agrees with the outcome 
of ChPT at tree level, only a subclass of chiral loop corrections 
(the ``unitarity class'' of rescattering diagrams) is effectively summed to infinite 
order in our model amplitude $f_{0+}$. Moreover, the loop function of Eq.~(\ref{eq:Gloop}) 
does not manifestly satisfy the rules of chiral power counting, and presumably the regulator 
scales $\alpha_{jb}$ will have some quark-mass dependence. Consequently, one should not expect 
that our fit results for the LECs will agree exactly with those found in a strict chiral-perturbative 
treatment. They should rather be considered as effective model parameters, which would only 
agree with order-of-magnitude estimates from ChPT results. This is a price one has to pay 
if one wants to extend the description of data beyond the limits of chiral effective field theory, 
especially in the resonance region.

Consider, for example, the LECs $b_{0}$, $b_{D}$ and $b_{F}$ that can be obtained from fits 
to baryon masses and sigma terms. There, one usually employs a scheme where the loop functions 
obey the chiral power counting, so that loop contributions to the baryon masses start 
at $\mathcal{O}(m_{q}^{3/2})$ (or $\mathcal{O}(M^3)$ in the meson masses). If this is not done 
and one uses a loop function that does not obey the power counting, one should expect 
that different values for the three $b$-LECs will arise. However, it turns out that 
the $b_{D}$ value is rather insensitive to such a shift which, in the case of $b_{D}$, 
comes with a numerically small prefactor $\sim 3F^2-D^2$. Thus, it seems legitimate to take 
$b_{D}$ as an input from chiral analyses of baryon masses, which typically result 
in small values $b_{D}\approx 0.1\,\mathrm{GeV}^{-1}$ \cite{Borasoy:1996bx,Bruns:2012eh}. 
On the other hand, $b_{0}$ and $b_{F}$ should be treated as free parameters since their 
loop-function renormalisations cannot be neglected. Similar considerations apply 
to the other groups of LECs.

It should be stressed that similar caveats apply when comparing our results for the fitted 
parameters with those of other non-perturbative models. For example, even though we use 
exactly the same effective Lagrangian as \cite{Borasoy:2002mt}, there are some notable 
differences in the full amplitude. One must keep in mind that, employing a non-perturbative 
framework, the resulting amplitude is not fixed by the effective Lagrangian, as would be the case
in a strictly perturbative approach. The most relevant difference is the treatment 
of the loop functions. While in \cite{Borasoy:2002mt} dimensional regularization is used, 
treating the associated renormalization scales $\mu$ as fit parameters, we employ 
the loop function given in our Eq.~(\ref{eq:Gloop}) above. As noted in \cite{Borasoy:2002mt} 
below their Eq.~(19), such a choice is completely legitimate. However, one cannot expect 
the same numerical values for the fitted LECs resulting in the different approaches 
(in particular because the loop functions do not satisfy the 'naive' power-counting rules of ChPT). 
Moreover, since the potentials are truncated after $\mathcal{O}(p^2)$ and the loop graphs 
are summed to infinite order, the energy dependence of the model amplitudes is
not expected to be exactly the same. Do also note that the $u$-channel Born terms 
were simply omitted in the meson-baryon scattering amplitude constructed in \cite{Borasoy:2002mt}, 
even though their s-wave projections are in general not suppressed with respect to those 
of the $s$-channel Born terms at low energies. Finally, it is not the aim of our article 
to perform a fit to meson photo- and electroproduction, as done e.g. in \cite{Borasoy:2002mt} - 
clearly, one should anticipate some changes of our fitted LECs in such an extended analysis, 
even though some parameters might be available in the construction of an electroproduction model.

Obviously, the LECs found in the present work should also deviate from those reported in previous 
analyses that did not include the $\eta'$ meson (or, from the point of view of the flavor basis, 
a heavy singlet meson $\eta_{0}$) as an explicit degree of freedom. The impact of this additional 
degree of freedom on our fits, and the relation of our results to previous works, will be discussed 
in the next sections.

\section{Experimental data and fitting procedure}
\label{sec:fits}

In general, theoretical approaches based on chiral Lagrangians are assumed to work well at low energies, 
for small momenta of the interacting particles. However, in this article we aim at a simultaneous 
description of the $\eta N$ and $\eta' N$ systems with the latter channel opening at quite large 
energy $E_{th}(\eta'N)=1897$ MeV. Even if we limit ourselves to energies close to either the $\eta N$ 
or $\eta' N$ threshold, the lowest $\pi N$ channel will operate at relativistic energies. 
Still, it is worth testing if the approach can be used to describe effectively the experimental data 
in such a broad interval of energies, spanning from the $\pi N$ threshold, $E_{th}(\pi N)=1077$ MeV, 
up to about 2~GeV. 

The model parameters are fitted to:
\begin{itemize}
\item $\pi N \longrightarrow \pi N$ amplitudes for the $S_{11}$ and $S_{31}$ partial waves taken 
from the SAID database \cite{SAID:2018}, that cover the energy interval from 1095 MeV to 1600 MeV. 
Following the treatment of these data presented in \cite{Bruns:2010sv, Mai:2012PhD} and \cite{Cieply:2013sya} 
we assume a semiuniform absolute variation of the SAID amplitudes and set it to $0.005$ for energies 
below 1228 MeV, and to $0.03$ for energies above 1228 MeV. All single energies data up to 1600 MeV 
(30 data points in total) are included for the real and imaginary parts of the $S_{11}$ amplitudes. 
The data on the $S_{31}$ amplitudes are considered only up to $1450$ MeV (21 data points) to avoid 
the impact of the $\Delta(1620)$ resonance.
\item selected $\pi^{-} p \rightarrow \eta n$ reaction cross section data in the energy region 
from 1500 up to 1600 MeV (10 data points). At the lowest energies up to 1525 MeV we include exclusively 
the modern data measured by the Crystal Ball collaboration \cite{Prakhov:2005qb} and complement 
them by the older bubble chamber data taken from \cite{Bulos:1970zk, Richards:1970cy, Feltesse:1975nz} 
to cover the higher energies as well.
\item $\pi^{-} p \rightarrow K^{0} \Lambda$ reaction cross section data in the energy region 
up to 1750 MeV (50 data points) \cite{Baldini:1988th}.
\item $\pi^{-} p \rightarrow \eta' n$ reaction cross section data for the lowest energies 
below 2~GeV (just 4 data points) \cite{Baldini:1988th}.
\end{itemize}

When making a comparison with the SAID data one has to multiply the $\pi N$ amplitudes generated 
by our model by the magnitude of the c.m.~momentum,
\beq
q_{\pi N}(s) [f_{\ell\pm}^{I}(s)]_{\pi N,\pi N} = T^{I}_{\ell\pm}(s) \; ,
\eeq{eq:qF}
since the SAID amplitudes $T^{I}_{\ell\pm}$ are dimensionless and normalized as
\beq
T^{I}_{\ell\pm} = e^{{\rm i} \delta_{\ell\pm}^{I}} \sin{\delta_{\ell\pm}^{I}} \;\;\; ,
\eeq{} 
where $\delta_{\ell\pm}^{I}$ denotes the phase shift in the $(2I,2J=2\ell\pm 1)$ wave. In general, the phase shift 
is a complex quantity when inelasticity is considered. For the elastic amplitude one can write
\beq{}
2{\rm i}\: T^{I}_{\ell\pm}  = e^{2{\rm i} (\mathrm{Re}\, \delta_{\ell\pm}^{I}+{\rm i}\mathrm{Im}\, \delta_{\ell\pm}^{I})} - 1 
= \eta_{\ell\pm}^{I} e^{2{\rm i} \mathrm{Re}\, \delta_{\ell\pm}^{I}} - 1 \;\;\; ,
\eeq{}
where we introduced the inelasticity factor $\eta_{\ell\pm}^{I} = \exp (-2\,\mathrm{Im}\, \delta_{\ell\pm}^{I})$. 
If needed, the real $\pi N$ phase shifts $\mathrm{Re}\, \delta_{\ell\pm}^{I}$ and inelasticities $\eta_{\ell\pm}^{I}$ 
are also provided by the SAID database. 

At this point we would also like to remind the reader that our approach is restricted to two-body 
interactions of pseudoscalar mesons with the basic SU(3) baryon octet. In reality, any other 
open channel not included in our approach does contribute to the inelasticities reported in the SAID 
database. At energies around the $\eta N$ threshold the $\pi\pi N$ channel already contributes 
to the total inelastic cross section for the $\pi N$-induced reactions,
\beq
\sigma_{r} \approx \frac{\pi}{q_{\pi N}^{2}} (1 - \eta_{0+}^{2})\,,
\eeq{eq:sigr}
where we neglect the influence of higher partial waves in the vicinity of the $\eta N$ threshold,
where the $N^{*}(1535)$ resonance dominates.
Comparing the reaction cross section calculated from the inelasticity reported in the SAID 
database with the maximum of the experimental $\pi^{-}p \rightarrow \eta n$ cross section 
one finds a difference of about 20\% at the peak energy.
One can effectively compensate 
for the missing $\pi \pi N$ channel by enhancing the calculated $\eta N$ cross sections. 
Thus, the calculated $\eta N$ cross section $\sigma_{I=1/2}$ is matched 
to the experimental one by using a relation
\beq
\sigma(\pi^{-}p \rightarrow \eta n) = \frac{2}{3} \sigma_{I=1/2}(\pi N \rightarrow \eta N) /1.2\,.
\eeq{eq:pipiN}
The estimate used here,
\beq
\epsilon_r(\sqrt{s}) := [1-\eta^{2}_{\mathrm{SAID}}(\sqrt{s})]/[1-\eta^{2}_{0+}(\sqrt{s})] \approx 1.2 \, ,
\eeq{eq:sr1}
works reasonably well in most part of the $N^{*}(1535)$ resonance region. However, $\epsilon_r$ must obviously diverge
at the $\eta N$ threshold, simply because our $\eta_{0+}=1$ below this threshold, while $\eta_{\mathrm{SAID}}<1$ there, due to
the presence of the $\pi\pi N$ channel in the SAID treatment. To account for this behaviour, we add a pole term in the
parameterization of $\epsilon_r$,
\beq
\epsilon_{r}^{\mathrm{eff}}(\sqrt{s}) := a/(\sqrt{s}-M_{\eta}-m_N) + b \, ,
\eeq{eq:sr2}
which (for suitably chosen parameters $a$ and $b$) describes quite well the energy dependence 
of the ratio $\epsilon_r$ in the whole interval from the $\eta N$ threshold up to the $K\Lambda$ threshold.
On the other hand, a factor $\epsilon_r\gg1$ just means that the $\pi\pi N$ channel,
which is not explicitly treated in our model amplitude, dominates the inelasticity close to the $\eta N$ threshold.
In order to avoid a strong dependence of our results on the chosen setting of $\epsilon_{r}(\sqrt{s})$,
and on the SAID parameterization in this region, we omit the first three data points for $\pi N\rightarrow\eta N$ 
above the $\eta N$ threshold (at energies $\sqrt{s} < 1500$ MeV) from our fitted data set.
The approach corresponding to the effective setting provided by Eqs.~(\ref{eq:pipiN}),~(\ref{eq:sr1}) 
was already used in \cite{Cieply:2013sya}, and is in agreement with observations made in Ref.~\cite{Gasparyan:2003fp}. 
The modifications that occur upon the use of the more sophisticated energy dependent $\epsilon_{r}^{\mathrm{eff}}(\sqrt{s})$ 
factor of Eq.~(\ref{eq:sr2}) will be discussed in the next section.

When performing the fits we use the MINUIT routine from CERNLIB to minimize the $\chi^{2}$ 
per degree of freedom defined as
\beq
\chi^{2}/dof = \frac{\sum_{i}N_{i}}{N_{obs}(\sum_{i}N_{i}-N_{par})} \sum_{i}\frac{\chi^{2}_{i}}{N_{i}} \; ,
\eeq{eq:chi2}
where $N_{par}$ is the number of fitted parameters, $N_{obs}$ is a number of observables, $N_{i}$ 
is the number of data points for an $i$-th observable, and $\chi^{2}_{i}$ stands for the total $\chi^{2}$ 
computed for the observable. Eq.~(\ref{eq:chi2}) guarantees an equal weight of the fitted data from 
various processes (i.e. for different observables).

Our chirally motivated approach contains a large number of parameters and it is essential 
to fix some of them to already established values. By doing so we reduce the number of degrees of freedom 
which in turn provides a better control over the fitting procedure. First of all it seems natural to adopt 
the following:
\begin{itemize}
\item Meson decay constants $f_{\pi}=92.4$ MeV, $f_{K}=110.0$ MeV, $f_{\eta}=118.8$ MeV as derived 
in \cite{Gasser:1984gg} and fine-tuned in fits of the $\bar{K}N$ related data \cite{Ikeda:2012au, Cieply:2011nq}.
\item The Born term couplings $F=0.46$ and $D=0.80$ as extracted in analysis of hyperon decays \cite{Ratcliffe:1998su}.
\item $c_{0,D,F} = 0$, $d_{5-7} = 0$ assumed to keep the number of fitted LECs at a reasonable level.
\item $b_D = 0.1$ GeV$^{-1}$, about the average value from various fits and estimates available in the literature. 
As we already mentioned above, unlike $b_0$ and $b_F$, the $b_D$ coupling is numerically not sensitive 
to the "renormalization" due to loop function contributions.
\item $D_s$ set to be from the interval $\langle -0.6,-0.2\,\rangle$, motivated by fits of the $\eta$ and $\eta'$ 
photoproduction and electroproduction data \cite{Borasoy:2002mt}, and compatible with the estimates for 
the $g_{\eta'NN}$ coupling \cite{Sibirtsev:2003ng}. After finding the $\chi^{2}$ minimum the $D_s$ value
is fine-tuned in the next step.
\end{itemize}

The inclusion of the $\eta'N$ channel in the  meson-baryon interactions means that we have five more 
parameters ($f_{\eta'}$, $\alpha_{\eta'N}$, $D_s$, $w_{s}$ and the pseudoscalar meson singlet-octet 
mixing angle $\theta$) when compared with the models that disregard the $\eta_{0} - \eta_{8}$ mixing. 
The value of $\theta = -15.5^\circ$ is well established from the analysis of the $\eta$ and $\eta'$ 
decays \cite{Bramon:1997va} but the remaining four parameters seem far too many to be reliably constrained 
by the $\pi^{-} p \rightarrow \eta' n$ cross section data that are rather scarce and not very precise 
at energies close to the $\eta'n$ threshold. Still, they represent the only data set related directly 
to the $\eta'N$ sector. For this reason we also assume that the $\eta'$ meson decay constant 
has the same value as the one adopted for the $\eta$, in accordance with the analysis 
of two photon decays of $\pi^{0}$, $\eta$ and $\eta'$ \cite{Borasoy:2003yb}.
Further, we perform the fits for a fixed value of $D_s$ which helps to stabilize the computer 
performance of the routine used to search for the $\chi^{2}$ minima. Thus, only the inverse range 
$\alpha_{\eta'N}$ and the parameter $w_{s}$ are left free to be determined from the interplay of 
the $\pi^{-} p \rightarrow \eta n$ and $\eta'n$ cross sections data.

To summarize, we are left with 12 free parameters to be determined in the fits and these consist of:
5 inverse ranges $\alpha_{jb}$, the $b_0$ and $b_F$ couplings, 4 $d$-couplings $d_{1-4}$, and $w_{s}$. 
Moreover, we have imposed additional restrictions on some of these parameters to keep them within 
reasonable limits during the fitting procedure. In particular, we have used 
$\alpha_{jb} \!\in\! \langle 400, 1200 \,\rangle$ MeV 
and $w_{s} \!\in\! \langle -0.3, 0.3 \,\rangle$. The restrictions imposed on the $w_{s}$ parameter 
may require additional comments as its value is not constrained by any theoretical nor experimental 
predictions. To our knowledge the pertinent contribution of the $w_{s}$ term to meson-baryon interactions 
was previously determined only in the low energy $\eta$ and $\eta'$ photoproduction fits 
performed in \cite{Borasoy:2002mt} where a small value of $w_{s} = -0.0125$ was reported (note, 
though, that this coupling appears with a large prefactor in the $\eta'N$ scattering amplitude).
In contrast, we have observed that our fits favour a relatively large negative value of $w_{s} \approx -1$. 
However, such a large value of $w_{s}$ leads to so strong attraction in the $\eta'N$ state that manifests 
as an appearance of an unphysical narrow resonance with a pole located either on the physical Riemann sheet (RS)
or very close to the real axis on the adjacent RS at energy between the $\pi N$ and $\eta N$
thresholds. The large $C_{\eta'N,\eta'N}$ coupling due to $w_{s} \approx -1$ that causes the effect 
is also several times bigger than other non-zero diagonal couplings in the same channel.
Thus, it seems natural to avoid this hindrance by enforcing 
the $w_{s}$ value reasonably small (of ``natural size'' compared to the other terms). 

In general, the relatively large parameter space complicates the search for local $\chi^{2}$ minima 
by appearance of solutions that suffer from unphysical resonant states represented by poles either 
on the physical RS or on the "second RS" (connected directly with the physical one), 
quite close to the real axis where no such state should exist. Thus, the $S$-matrix for each solution 
(combination of the fitted parameters) should be checked to be free of such spurious states. 
To ensure this, we have searched for poles on the physical and second Riemann sheets 
in a broad region of complex energies with imaginary parts as far as 150 MeV away from the real axis. 
If any unphysical poles were found, the $\chi^{2}$ minimum was excluded even if this meant choosing another 
local minimum with a worse (higher) $\chi^{2}$ value. 

The fits were performed under several different conditions for varied fixed values of $D_s$ 
(usually $D_s = -0.2, \;-0.4, \;-0.6$). When an acceptable local minimum was found the $D_s$ value 
was tuned to achieve the best possible $\chi^{2}$. Here we report the best solutions found 
under the following fitting scenarios:
\begin{itemize}
\item[{\bf model A}] global fit to the experimental data with the $\pi\pi N$ channel effectively accounted 
  for by enhancing the fitted $\eta N$ cross sections by an energy dependent factor $\epsilon_{r}^{\mathrm{eff}}$
  adjusted to provide the $\pi N$ inelasticities from the SAID database \cite{SAID:2018},
\item[{\bf model B}] global fit to the experimental data with the $\pi\pi N$ channel effectively 
  accounted for by enhancing the fitted $\eta N$ cross sections by an effective factor of $1.2$,
\item[{\bf model C}] "low energy" fit of experimental data restricted to energies $\sqrt{s} \leq 1600$ MeV, 
  with no $\eta_{0} - \eta_{8}$ mixing ($\theta = 0$), the $\eta'N$ channel decoupled, 
  and the fitted $\eta N$ cross sections enhanced by an effective factor of $1.2$,
\item[{\bf model D}] global fit performed disregarding the impact (inelasticity) of any reactions 
  not accounted for within our meson-baryon channel space,
\item[{\bf model E}] global fit to the experimental data with the $\eta_{0} - \eta_{8}$ mixing 
  switched off.
\end{itemize}

First of all we performed a fit (model A) to all experimental data specified above that cover 
a very broad interval of energies from the $\pi N$ threshold $E_{\pi N} = 1077$ MeV 
up to almost $2000$ MeV involved in the $\eta'n$ cross sections. For this fit we used an energy 
dependent factor $\epsilon_{r}^{\mathrm{eff}}(\sqrt{s})$, Eq.~(\ref{eq:sr2}), to enhance the fitted $\eta N$ 
cross sections and compensate the absence of some channels in our model.\footnote{ 
The two parameters $a$ and $b$ of Eq.~(\ref{eq:sr2}) were adjusted selfconsistently
by matching the inelasticities (multiplied by the $\epsilon_r$ factor) of the computed 
$\pi N$ amplitudes to those provided in \cite{SAID:2018}. The exact values determined 
in the A model fit are $a=1.7448$ MeV and $b=1.0972$.} 
The model B represents a fit to the same set of the data for the $\epsilon_r$ 
factor fixed at the $1.2$ value which provides quite realistic approximation of the factor 
at energies from about $1520$ MeV on. The fit provided by model C was performed with 
the experimental data restricted to low energies including only the $\pi N$ amplitudes 
and the $\eta N$ cross sections data. Since the $\eta'N$ channel is not involved 
at the low energies we have also decoupled it completely and disregarded 
the $\eta_{0} - \eta_{8}$ mixing in the model C scenario. This, and adopting the $\epsilon_r=1.2$ value, 
makes the model directly comparable with the one presented in \cite{Cieply:2013sya}. 
Finally, the models D and E are presented to demonstrate the impact of omitting completely 
the effective treatment of the $\pi\pi N$ channel (by setting $\epsilon_r = 1$) and of switching 
off the $\eta_{0} - \eta_{8}$ mixing, respectively.

\begin{table}[h]
\centering
\caption{The fit results and parameters of our models. The inverse ranges $\alpha_{jb}$ are in MeV, 
the NLO couplings $b$ and $d$ in GeV$^{-1}$.}
\begin{tabular}{c|ccccc}
model                &   A    &   B    &   C    &   D     &   E     \\ \hline
$\chi^{2}/dof$       &  2.21  &  2.12  &  0.78  &  2.44   &  2.04   \\ 
$\alpha_{\pi N}$     &   596  &   629  &   581  &   569   &   668   \\
$\alpha_{\eta N}$    &   959  &   959  &   953  &   966   &   973   \\
$\alpha_{K\Lambda}$  &  1188  &  1200  &   788  &  1172   &  1200   \\
$\alpha_{K \Sigma}$  &   443  &   447  &   400  &   434   &   454   \\
$\alpha_{\eta' N}$   &   911  &   916  &   ---  &   923   &  1200   \\
$b_0$                & -0.452 & -0.415 & -0.673 & -0.488  & -0.368  \\
$b_F$                & -0.049 & -0.028 &  0.184 & -0.077  & -0.002  \\
$d_1$                & -1.648 & -1.643 &  0.630 & -1.654  & -1.638  \\
$d_2$                &  0.574 &  0.569 &  0.161 &  0.572  &  0.696  \\
$d_3$                &  1.190 &  1.263 &  3.547 &  1.115  &  1.252  \\
$d_4$                & -0.332 & -0.329 & -1.302 & -0.336  & -0.400  \\
$w_{s}$              & -0.038 &  0.011 &   ---  & -0.110  & -0.236  \\
$D_s$                & -0.28  & -0.27  &   ---  & -0.33   & -0.29  
\end{tabular}
\label{tab:fits}
\end{table}

The parameters of our models are provided in Table~\ref{tab:fits} where we show the resulting $\chi^{2}/dof$ 
as well. We also note that the number of fitted parameters is equal to 12 for all models excluding model C 
where $N_{par}=10$ since $\alpha_{\eta'N}$ and $w_{s}$ are not used in the latter. As the sets of fitted experimental 
data also vary for the listed models their $\chi^{2}/dof$ values are suitable for measuring the quality 
of the fits but may not be directly comparable among themselves. The A model fit provides a satisfactory 
reproduction of the data from the whole energy region, though the quality of the {\it "low energy"} 
C model fit is obviously much better when one considers only data in the pertinent energy interval. It should 
also be noted that the model parameters vary only moderately when the fit is performed for various 
settings of the effective inelasticity treatment that represents the only difference between models A, B, and D.
The exception here is a large negative value of $w_s$ for the model D, apparently not compatible 
with its earlier estimate in fits of the $\eta$ and $\eta'$ photoproduction data \cite{Borasoy:2002mt}.
An even larger $w_s$ value is obtained in the E model fit. As the other parameters do not deviate 
much from those found in the A (or B and D) model fit it appears that the difference in the $w_s$ value 
is solely responsible for compensating the effects due to switching off the $\eta_{0} - \eta_{8}$ mixing.

Finally, we remark that the NLO LECs of our models, the $d$-couplings, are much larger than those 
reported in the $\eta$ and $\eta'$ photo- and electro-production analysis \cite{Borasoy:2002mt}. 
Although this feature raises concerns about the convergence of the perturbative treatment it seems 
difficult to avoid considering the broad interval of energies covered in our work. It is also known that 
the subleading terms may be enlarged due to the presence of nearby resonances as noted e.g.~in the chiral 
expansion of the two-nucleon forces \cite{Epelbaum:2008ga} or in some analyses of the $\bar{K}N$ 
data~\cite{Cieply:2016jby}. Thus, relatively large NLO terms are not that uncommon and cannot be ruled out.

\section{Results}
\label{sec:results}

\subsection{Data reproduction}
\label{sec:fitres}

\begin{figure}[htb]    
\centering
\includegraphics[width=0.48\textwidth]{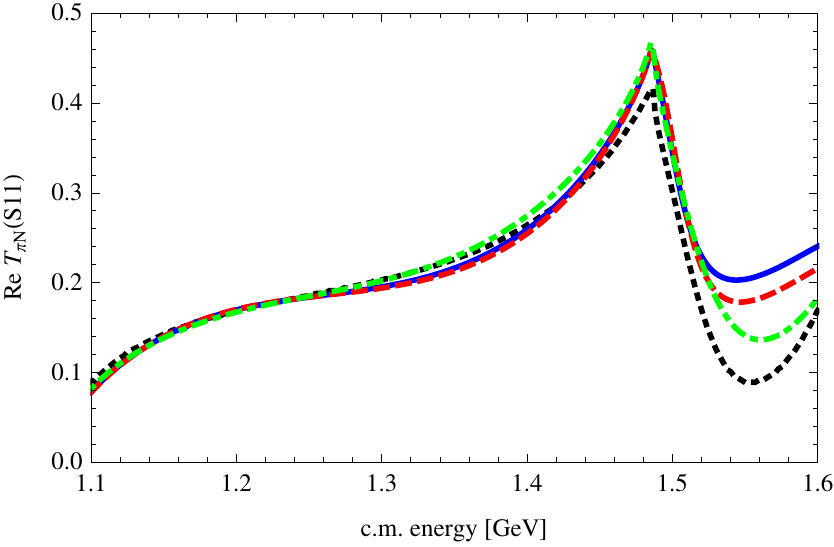}\hspace*{2mm}
\includegraphics[width=0.48\textwidth]{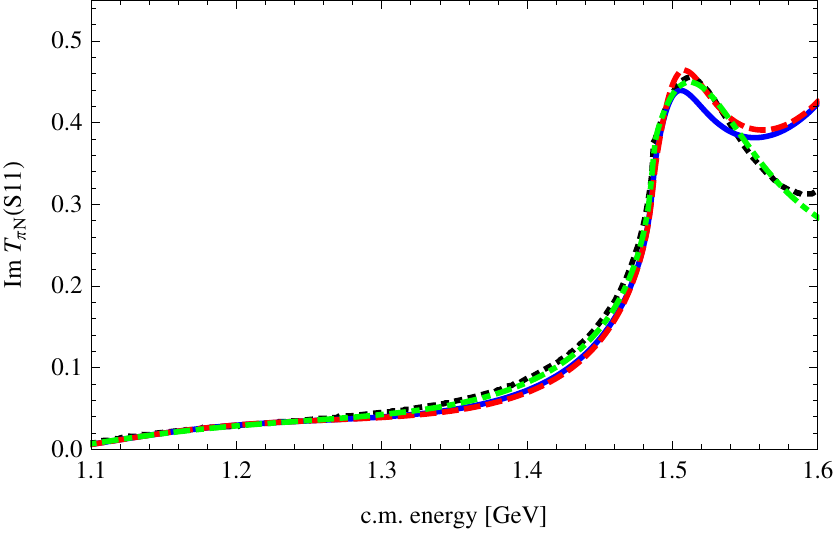} \\
\includegraphics[width=0.48\textwidth]{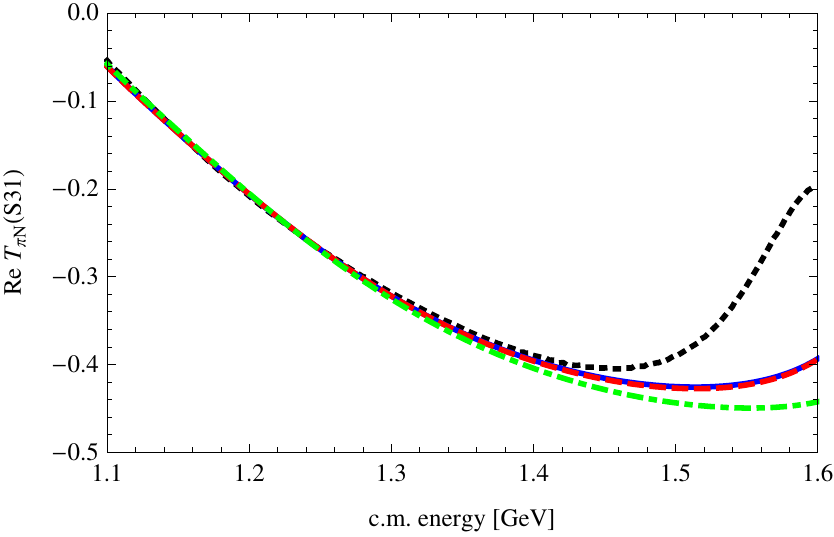}\hspace*{2mm}
\includegraphics[width=0.48\textwidth]{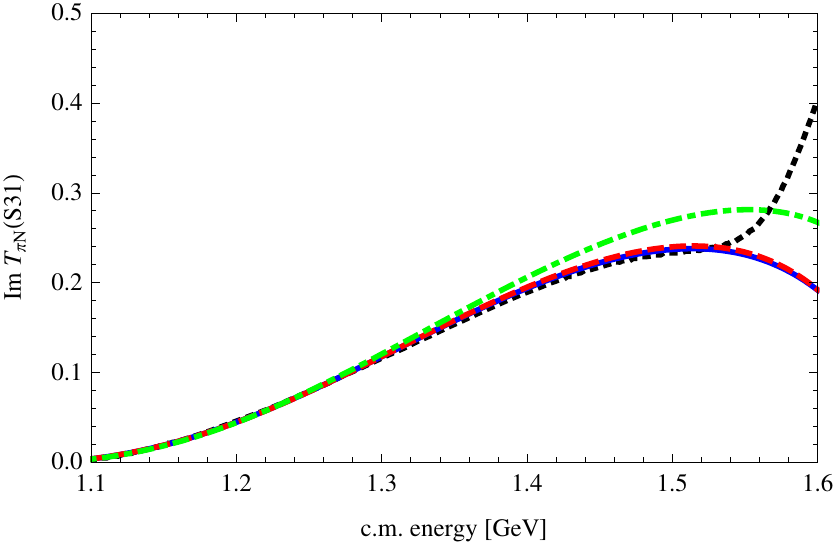}
\caption{The real (left panels) and imaginary (right panels) parts of the $T_{\pi N}(S11)$ (top panels) and 
$T_{\pi N}(S31)$ (bottom panels) amplitudes generated by our models A (blue continuous lines), 
B (red dashed lines) and C (green dot-dashed lines). The dotted black lines represent the SAID partial 
wave solution \cite{SAID:2018}.
}
\label{fig:piNamps}
\end{figure}

In Fig.~\ref{fig:piNamps} we present the dimensionless amplitudes $T_{\pi N}(S11)=T^{1/2}_{0+}$ 
and $T_{\pi N}(S31)=T^{3/2}_{0+}$ defined by Eq.~(\ref{eq:qF}) and generated by our models A, B and C. 
It is remarkable how well all three models reproduce the SAID amplitudes \cite{SAID:2018} 
over a very broad interval of energies up to about 1500 MeV. Naturally, the {\it low energy fit} 
represented by model C is doing well in the $S_{11}$ sector even in the {\it dip region} of energies 
from the $\eta N$ threshold up to $1600$ MeV.
Of course, the model predictions for the $S_{31}$ amplitude start to deviate earlier, around 
1450 MeV, because of the presence of the $\Delta(1620)$ resonance that is not accounted for 
in our model. A good reproduction of the $I=3/2$ amplitude up to 1450 MeV also justifies our choice 
of this energy as the upper limit beyond which the $I=3/2$ data are disregarded in the fits. 
The reproduction of the $S_{31}$ partial wave 
at higher energies could be improved by introducing explicitly the $\Delta(1620)$ resonance 
in the model. However, this would go beyond the scope of the current approach incorporating only 
two-body meson-baryon channels. It also seems that the inclusion of the $\Delta(1620)$ would 
just add the resonance on top of the $\pi N - K\Sigma$ coupled channels background seen 
in Fig.~\ref{fig:piNamps} and hardly affect the fitted parameters of our models.

\begin{figure}[h]    
\centering
\includegraphics[width=0.48\textwidth]{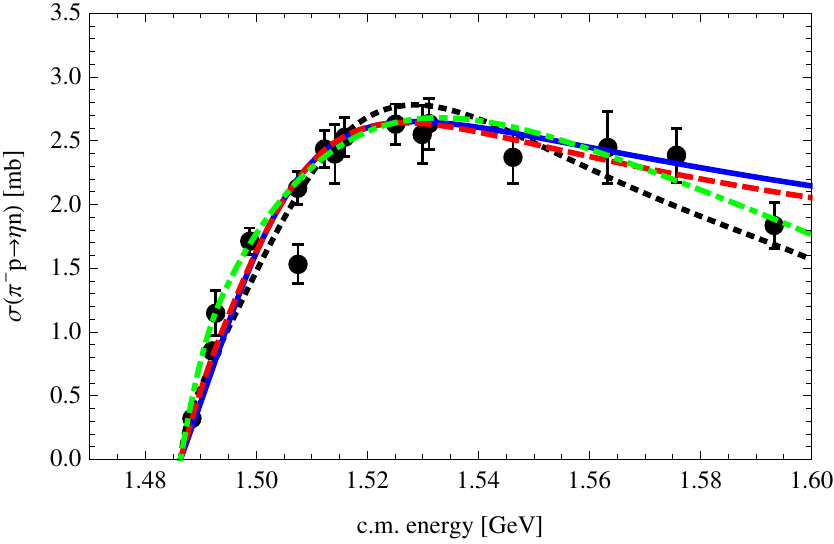}\hspace*{2mm}
\includegraphics[width=0.48\textwidth]{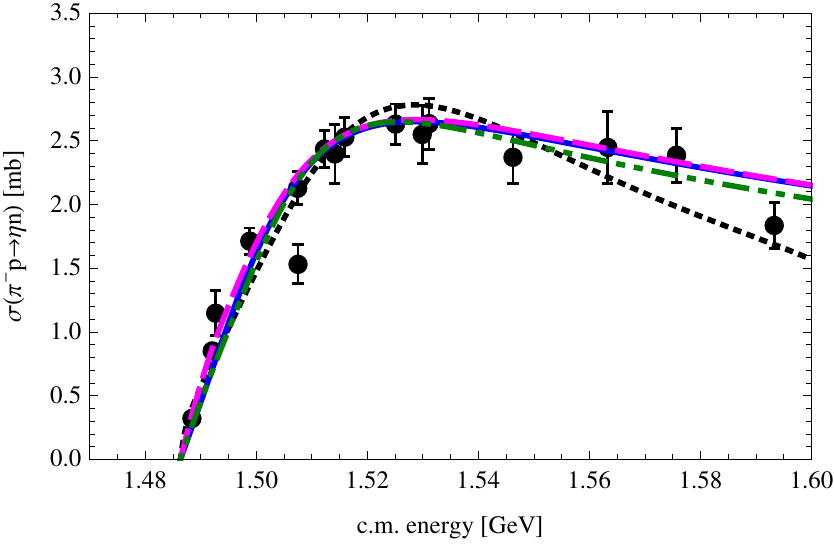} \\
\caption{A comparison of various model predictions for the $\pi^{-}p \rightarrow \eta n$ cross section. 
Left panel: The results obtained with our models A (blue continuous line), B (red dashed line) 
and C (green dot-dashed line) are plotted together with the experimental data. The CS model 
predictions \cite{Cieply:2013sya} represented by the black dotted line are shown for a comparison as well. 
Right panel: The same for models A (blue continuous line), D (magenta long-dashed line) 
and E (dark green dot-dot-dashed line).}
\label{fig:etaNxs}
\end{figure}

In the left panel of Fig.~\ref{fig:etaNxs} we show how our models A (blue continuous line), 
B (red dashed line) and C (green dot-dashed line) reproduce the $\eta n$ production cross section data. 
Our results are plotted in comparison with those taken from Ref.~\cite{Cieply:2013sya} and visualized 
by the dotted black line. All our three models reproduce the data about equally well and provide 
higher $\eta n$ cross sections than the CS model (the NLO30$_{\eta}$ model from \cite{Cieply:2013sya}) 
at energies above the $N^{*}(1535)$ resonance. The resonant peak also seems to be more pronounced 
in the CS model, though one cannot say that the data reproduction is worse with 
our current approach. The right panel of Fig.~\ref{fig:etaNxs} demonstrates the impact 
of accounting effectively for the $\pi\pi N$ channel (model D, magenta long-dashed line) 
or of switching off the $\eta_{0} - \eta_{8}$ mixing (model E, dark green dot-dot-dashed line). 
The D model reproduction of the $\eta n$ cross section is hard to 
distinguish in the figure from the A model predictions, though an overall 
$\chi^{2}/dof$ value is moderately worse in the case of the D model. We conclude that 
accounting for the $\pi\pi N$ (or any other not included in our approach) channel is appropriate 
but does not have significant impact on our results for the $\eta N$ cross sections. 

Setting the $\eta_{0} - \eta_{8}$ mixing angle to zero in the E model seems almost fully 
compensated by the larger (negative) value of the $w_s$ parameter obtained in the fit. 
It should be remembered that the $\eta'N$ channel remains coupled to the other channels even 
when the mixing is switched off which makes the model different from the C model scenario 
and allows for the compensation either due to $w_s$ or $D_s$ variations. In reality, the $D_s$ 
value appears to be quite stable in our fits. The right panel of Fig.~\ref{fig:etaNxs} demonstrates 
that the $\eta N$ cross sections calculated with the E model are only marginally different 
from those obtained with the A and D models. We have also checked what happens when 
the $\eta_{0} - \eta_{8}$ mixing is switched off without re-fitting the model parameters. 
Then the $\eta N$ cross sections would differ moderately from those seen in Fig.~\ref{fig:etaNxs}
with the $N^{*}(1535)$ resonance peak shifted to lower energies and being more pronounced,  
the latter effect resembling the results of Ref.~\cite{Cieply:2013sya}.  

\begin{figure}[h]    
\centering
\includegraphics[width=0.48\textwidth]{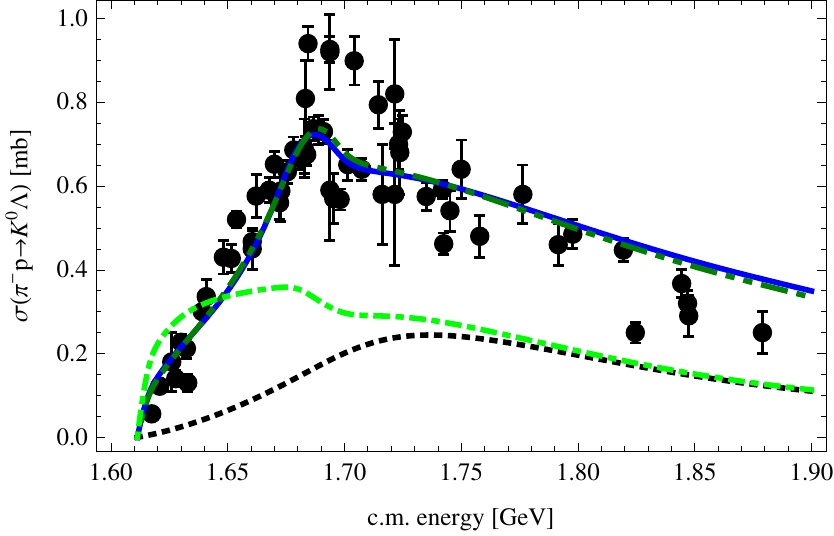}\hspace*{2mm}
\includegraphics[width=0.48\textwidth]{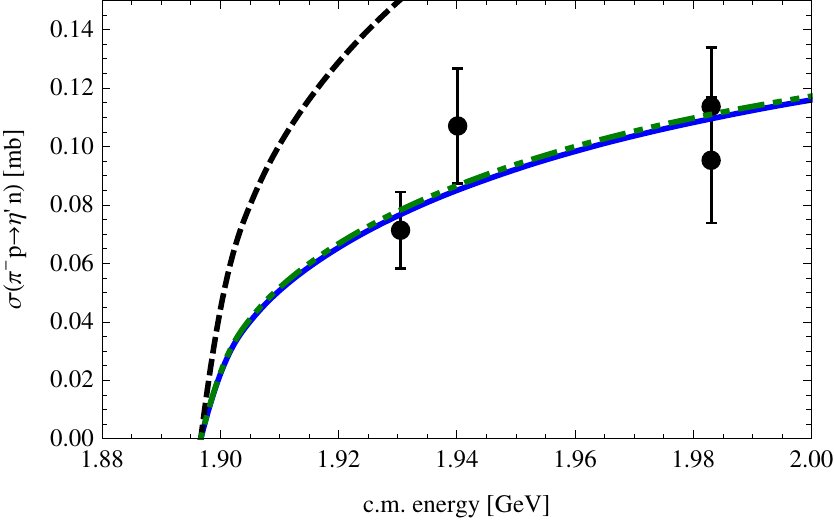} \\
\caption{A comparison of our model predictions for the $\pi^{-}p \rightarrow K^{0}\Lambda$ 
and $\pi^{-}p \rightarrow \eta'n$ cross sections. 
Left panel: The $K^{0}\Lambda$ results obtained with our A (blue continuous line), E (dark green dot-dot-dashed line) 
and C (green dot-dashed line) models are plotted together with the experimental data. The black dotted line 
visualizes the p-wave contribution provided by the Bonn-Gatchina analysis \cite{BonnGatchina:2014}.  
Right panel: The $\eta'n$ results for models A (blue continuous line) and E (dark green dot-dot-dashed line). 
The black dashed line shows the effect of switching-off the $\eta_{0} - \eta_{8}$ mixing while keeping 
the A model parameter set.}
\label{fig:HExs}
\end{figure}

In Fig.~\ref{fig:HExs} we show aside the model predictions for the $K^{0}\Lambda$ and $\eta'n$ 
cross sections. As there is a sizeable p-wave contribution to the $K^{0}\Lambda$ total cross sections 
we construct it from the p-wave amplitudes provided by the Bonn-Gatchina analysis \cite{BonnGatchina:2014} 
and add the p-wave cross sections to the s-wave ones generated by our models. The respective size 
of the s-wave and p-wave contributions can be seen in the figure where the Bonn-Gatchina p-wave 
cross-sections are visualized by the dotted black line. Our models A, B, D and E provide an equally 
good description of the $K^{0}\Lambda$ and $\eta'n$ experimental data with the calculated cross sections 
only marginally different in these models and practically indistinguishable in the figures. 
For this reason, to prevent an overlap of multiple lines, only cross sections generated by the A and E 
models are shown in Fig.~\ref{fig:HExs} to represent these four {\it equivalent} predictions.
The {\it low energy} C model added in the left panel of the figure clearly does not reproduce 
the $K^{0}\Lambda$ production data and starts to deviate from them already about 20 MeV above 
the reaction threshold. It comes as no surprise as the $K^{0}\Lambda$ cross sections data were excluded 
from fitting the model C parameters. The latter model also does not account for the $\eta'n$ data as 
this channel is purposely completely decoupled in this scenario. 

For comparison, in the right panel of Fig.~\ref{fig:HExs} we also demonstrate what happens when one 
simply switches-off the $\eta_{0} - \eta_{8}$ mixing while not altering the parameters of the A model.
As anticipated, the predicted $\eta'n$ cross sections then strongly deviate from those generated 
by the other models that reproduce nicely the experimental data. It also means that this large difference 
is completely compensated by re-fitting the parameters in the E model scenario. Interestingly, as one can
check in Table \ref{tab:fits}, only the $w_s$ parameter deviates significantly when the E model 
and A model parameter sets are compared. In other words, the mixing angle $\vartheta$ and the $w_s$ 
parameter appear to be closely correlated. Studying this effect, we found that especially the $\eta'N \rightarrow \eta'N$
potential is quite sensitive to small variations of the mixing angle when the other parameters are fixed 
at their fitted values. The singlet-singlet $w_{s}$ coupling has the biggest impact 
in this potential, and is able to provide compensation for variations of the mixing angle, 
while the other couplings are more tightly constrained by observables in other channels.

\subsection{$\eta N$ and $\eta'N$ amplitudes}
\label{sec:EtaN}

We begin our discussion of the model predictions for the $\eta N$ and $\eta'N$ amplitudes with 
a presentation of the computed scattering lengths given in Table \ref{tab:ath}. There, we also 
show the calculated $S_{11}$ part of the $\pi N$ scattering length. All our models are in nice 
agreement concerning the $S_{11}$ $\pi N$ scattering length and also reasonably compatible 
with the chiral prediction of $a_{\pi N}(S_{11}) = 0.140$ fm calculated at the tree level 
including the Born and contact terms at the ${\cal O}(p^{2})$ order \cite{Kaiser:2001hr} and 
adopting the A model LECs. The loop contributions are then responsible for any difference between 
the tree level estimate and the A model result provided in Table~\ref{tab:ath}. 

\begin{table}[h]
\centering
\caption{The $S_{11}$ scattering lengths (in fm) generated by our models 
for the $\pi N$, $\eta N$ and $\eta'N$ channels.}
\begin{tabular}{c|ccccc}
model         &      A        &       B       &       C       &       D       &       E       \\ \hline
$\pi N$       & ( 0.20, 0.00) & ( 0.20, 0.00) & ( 0.22, 0.00) & ( 0.21, 0.00) & ( 0.20, 0.00) \\ 
$\eta N$      & ( 1.05, 0.17) & ( 0.86, 0.13) & ( 0.73, 0.26) & ( 1.10, 0.12) & ( 0.85, 0.09) \\
$\eta'N$      & (-0.41, 0.04) & (-0.41, 0.04) &       ---     & (-0.41, 0.04) & (-0.29, 0.04) 
\end{tabular}
\label{tab:ath}
\end{table}

The $a_{\eta N}$ scattering length obtained with our {\it low energy} model C is in a good agrement 
with the previous estimates of $a_{\eta N} = (0.67 + {\rm i}\,0.20)$ fm and $(0.77 + {\rm i}\,0.22)$ fm 
provided under similar model settings in \cite{Cieply:2013sya} and \cite{Nieves:2001wt}, respectively.
The real part of the $a_{\eta N}$ scattering length in model A is much larger than in model C confirming  
the prediction of \cite{Bass:2005hn} that the $\eta_0$ component in the $\eta$ meson should increase 
the $\mathrm{Re}\, a_{\eta N}$ value. The model A prediction makes the $\eta N$ interaction even more attractive 
at the threshold than determined in the $K$-matrix analysis, where the value $a_{\eta N} = 0.91(6) + {\rm i}\,0.27(2)$ fm 
was found \cite{Green:2004tj}. It should also be noted that a sizeable $\eta N$ attraction increases the chance 
that $\eta$-nuclear bound states can be observed \cite{Cieply:2013sga}. In particular, the value of 
$\mathrm{Re}\, a_{\eta N} \approx 1$ fm might be a prerequisite for a formation of the $\eta$-$^{3}$He and 
$\eta$-$^{4}$He bound states~\cite{Barnea:2017epo}. We also mention that the imaginary part of the 
$a_{\eta N}$ value obtained with the A model just complies with the lower bound imposed by unitarity 
from the analysis of the experimental $\pi N \rightarrow \eta N$ cross sections, namely 
$\mathrm{Im}\, a_{\eta N} > 0.172 \pm 0.009$ fm \cite{Arndt:2005dg}. The model B and D predictions underestimate 
the $\mathrm{Im}\, a_{\eta N}$ value, apparently due to lacking the energy dependence provided near the $\eta N$ 
threshold in the effective treatment of inelasticities by means of the $\epsilon_{r}^{\mathrm{eff}}$ factor, 
Eq.~(\ref{eq:sr2}).

The $\eta'N$ scattering length predicted by our models is remarkably stable with the real part falling 
within the recent estimates derived from the final state interactions in the $pp \longrightarrow pp\eta'$ 
reaction measurement at COSY \cite{Czerwinski:2014yot},
\begin{displaymath}
\mathrm{Re}\, a_{\eta'N} = 0 \pm 0.43           \;{\rm fm,} \quad
\mathrm{Im}\, a_{\eta'N} = 0.37^{+0.40}_{-0.16} \;{\rm fm.}
\end{displaymath}
The imaginary part of $a_{\eta' N}$ generated by our models appears 
to be too small which we attribute to limitations of our approach, in particular to not including 
channels beyond the pseudoscalar meson-baryon ones.

\begin{figure}[h]    
\centering
\includegraphics[width=0.48\textwidth]{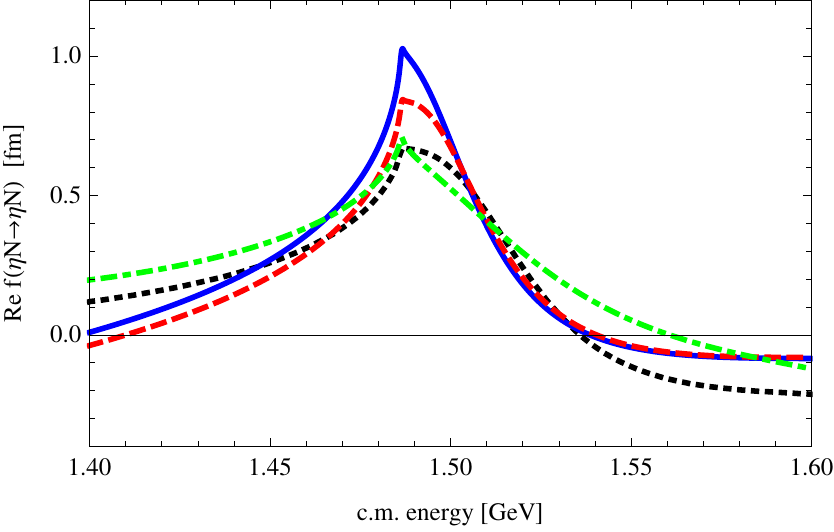}\hspace*{2mm}
\includegraphics[width=0.48\textwidth]{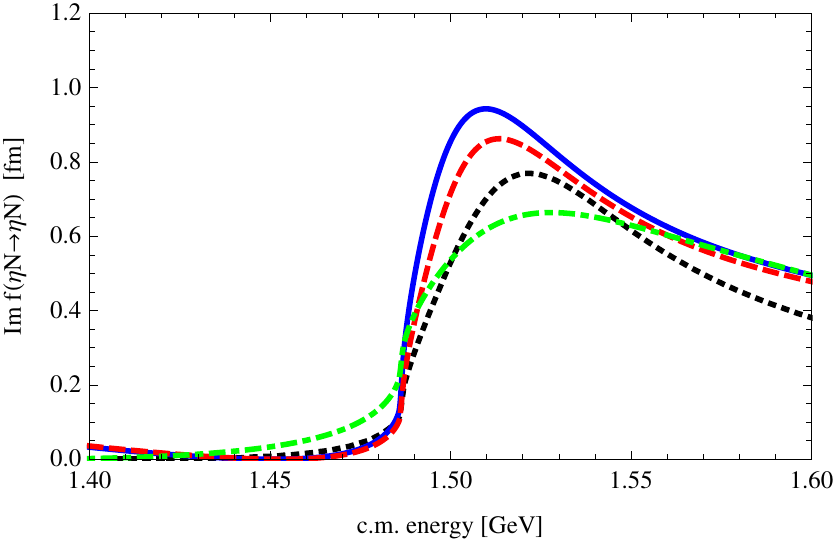} \\
\caption{Model predictions for the elastic $\eta N \rightarrow \eta N$ amplitude. The real (left panel) 
and imaginary (right panel) parts of the amplitude generated by our models A (blue continuous line), 
B (red dashed line) and C (green dot-dashed line) are shown in comparison with the CS model 
predictions \cite{Cieply:2013sya} visualized by the black dotted lines.}
\label{fig:AetaN}
\end{figure}

The energy dependence of the $\eta N$ elastic scattering amplitude is shown in Fig.~\ref{fig:AetaN} 
where our current predictions are compared with those from Ref.~\cite{Cieply:2013sga}. 
In the energy region above the $\eta N$ threshold the elastic amplitude is clearly dominated 
by the $N^{*}(1535)$ resonance, though the peak in the imaginary part of the $\eta N$ amplitude 
appears at about 20-30 MeV lower energy when compared with the nominal value. 
While at the $\eta N$ threshold our model C amplitude is in good agreement with the $a_{\eta N}$ 
value reported in \cite{Cieply:2013sga} one observes moderate variations of the amplitude 
especially at energies above the threshold. In particular, the $N^{*}(1535)$ peak manifested 
in the imaginary part of the amplitude is broader for the C model than the one for the CS model. 
We have checked that this difference can be attributed to the omission of the first three 
experimental data points for the $\eta N$ cross sections in our current fits. If these data were 
included in our C model fits we would get much better agreeement with the CS model for both,
the scattering length and the energy dependence of the $\eta N$ amplitude.

\begin{figure}[h]    
\centering
\includegraphics[width=0.48\textwidth]{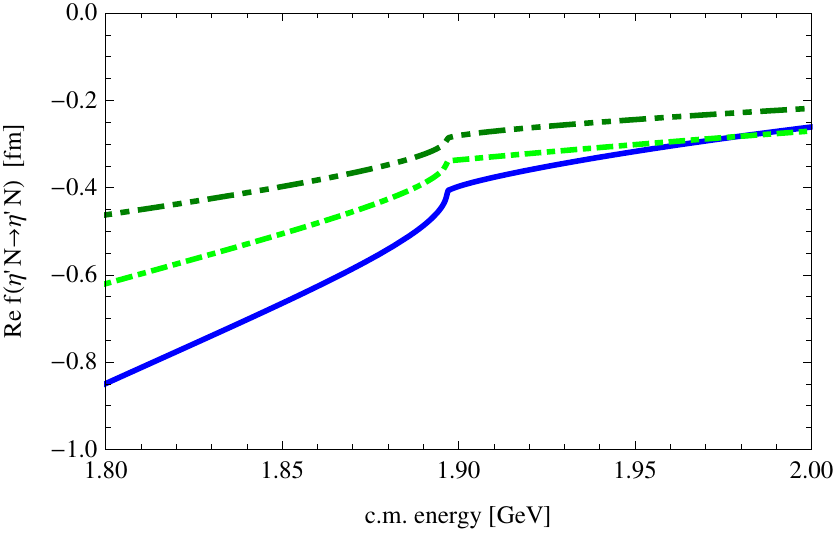}\hspace*{2mm}
\includegraphics[width=0.48\textwidth]{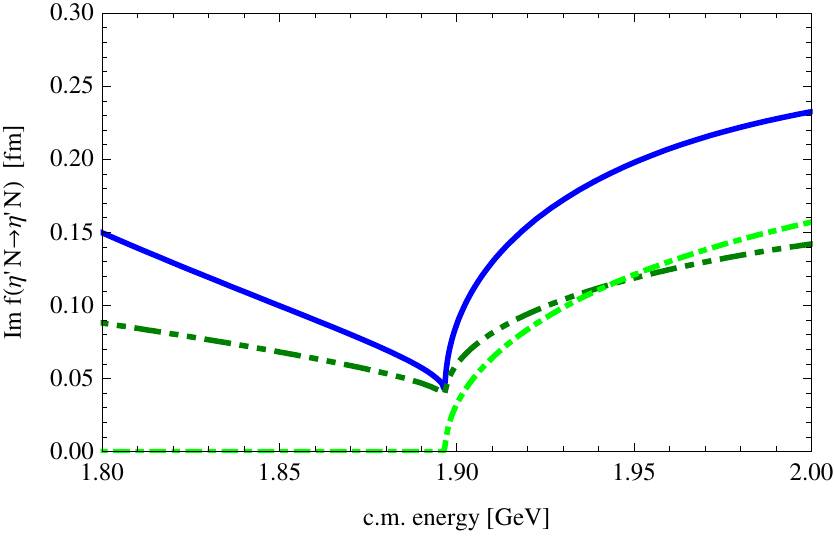} \\
\caption{Model predictions for the elastic $\eta'N \rightarrow \eta'N$ amplitude. The real (left panel) 
and imaginary (right panel) parts of the amplitude are presented as generated by our models 
A (blue continuous line), C (green dot-dashed line) and E (dark green dot-dot-dashed line).}
\label{fig:AetapN}
\end{figure}

Figure~\ref{fig:AetapN} shows the energy dependence of the $\eta' N$ elastic scattering amplitude 
for models A, C and E. The B and D models predictions coincide with those of model A and would 
overlap with the A model curves. It means that different approaches to the effective treatment 
of the inelasticity (various $\epsilon_r$ settings) do not have any impact on our results 
for both, the $\pi N \rightarrow \eta'N$ cross sections as well as for the $\eta' N$ scattering 
amplitude. On the other hand, energy dependence of the E model amplitude does differ moderately 
from the one generated by the A model despite both models providing practically the same predictions 
for the cross sections shown in Fig.~\ref{fig:HExs}. In other words, the low energy $\eta'N$ cross 
sections data do not provide sufficient restrictions to constrain the LECs related to 
the $\eta$-singlet sector. It is also difficult to determine better the $\eta' N$ amplitude  
due to insufficient coverage of the relevant energies by the available experimental data.

For all our models the real part of the $\eta' N$ elastic scattering amplitude remains negative 
in the whole energy region which relates to repulsive interaction. We have checked (for the $\eta'N$ 
threshold energy) that most of this repulsion is caused by large NLO $d$-terms with the (negative) 
$w_s$ term compensating partly to provide the $\eta'N$ scattering length appropriate to the fitted 
cross sections. 
Although there is no direct evidence concerning the character of the $\eta' N$ interaction 
there are indications that it should be attractive, e.g.~due to the $\eta'$ effective mass shift 
in nuclear medium deduced from the photoproduction experiments on nuclear targets \cite{Nanova:2016cyn}. 
Similar in-medium mass shifts were also predicted in theoretical calculations based on 
the Nambu-Jona-Lasinio model \cite{Nagahiro:2006dr} and on the linear sigma model \cite{Sakai:2013nba}. 
Finally, the $N^{*}(1895)$ resonance included recently in the Particle Data Group tables \cite{Tanabashi:2018oca} 
may also indicate an attractive $\eta' N$ interaction, which we will address in the following section. 
Therefore, our predictions of the repulsive $\eta' N$ elastic scattering amplitude may be taken 
with a grain of salt and viewed within the scope and limitations of the current approach. 

\subsection{Dynamically generated resonances}
\label{sec:poles}

The coupled channels chiral models restricted to pseusoscalar meson-baryon interactions are rather 
limited in their options to generate resonances. However, it was already demonstrated by several authors 
that the two most important states in the $S_{11}$ partial wave, the $N^{*}(1535)$ and $N^{*}(1650)$, 
can be reproduced reasonably well \cite{Kaiser:1995cy, Nieves:2001wt, Bruns:2010sv, Cieply:2013sya}. 
Both of them are generated dynamically within our model with strong couplings to the $K\Sigma$ channel. 
In Table \ref{tab:poles} we show the positions of the poles our models generate on two Riemann sheets 
that are connected with the physical region in the considered energy interval. The RS connected 
to the physical region by crossing the real axis between the $\eta N$ and $K \Lambda$ thresholds is denoted 
as [-,-,+,+,+] with the signs marking the signs of the imaginary parts of the meson-baryon c.m. momenta in all five 
coupled $I=1/2$ channels (unphysical for the $\pi N$ and $\eta N$ channels and physical for the remaining ones). 
Similarly, the RS connected with the physical region in between the $K\Lambda$ and $K\Sigma$ thresholds 
is denoted as [-,-,-,+,+].

\begin{table}[h]
\centering
\caption{The positions (complex energies in MeV) of the poles assigned to the $N^{*}(1535)$ 
and $N^{*}(1650)$ resonances.}
\begin{tabular}{cc|ccc}
  resonance    &      RS     &      A      &      B      &      C        \\ \hline
$N^{*}(1535)$  & [-,-,+,+,+] & (1490, -27) & (1493, -28) & (1487, -58)   \\ 
$N^{*}(1650)$  & [-,-,-,+,+] & (1709, -22) & (1710, -23) &     ---    
\end{tabular}
\label{tab:poles}
\end{table}

For the $N^{*}(1535)$ resonance the Particle Data Group (PDG) \cite{Tanabashi:2018oca} lists  
the real and imaginary parts of the pole energy in the intervals $\mathrm{Re}\, z \approx 1500 - 1520$ MeV 
and $-\mathrm{Im}\, z \approx 55 - 75$ MeV, respectively. All our three models listed in Table \ref{tab:poles} 
generate the pole about 10 MeV below the lower end of the PDG interval. The models A and B also 
provide too small decay width $\Gamma = -2\,\mathrm{Im}\, z$ while the C model has this value at about 
the lower end of the PDG estimates. The relatively small decay widths can be attributed 
to a lack of some channels in our approach which the $N^{*}(1535)$ resonance decays to. 
This seems reasonable as the $\pi N$ and $\eta N$ channels account to about $70-80\%$ 
of the total decay width \cite{Tanabashi:2018oca}. 

The PDG estimates for the $N^{*}(1650)$ resonance pole position 
are $\mathrm{Re}\, z \approx 1640 - 1670$ MeV and $\mathrm{Im}\, z \approx 50 - 85$ MeV. 
Our models A and B generate the pertinent pole at a higher energy, even above 
the $K \Sigma$ threshold. The C model is not expected to work well 
in the $N^{*}(1650)$ energy region as its parameters were not fitted to the 
$K\Lambda$ data and the model misses on the resonance pole completely. However, we noted 
that the C model still generates a pole located too far from the real axis, more 
than 200 MeV. Even for the A and B models, the $N^{*}(1650)$ pole is generated 
at too high energy, 60 MeV above the resonance nominal value. This is in contrast with 
a very good reproduction of the $K\Lambda$ production data by the A and B models 
as demonstrated in the left panel of Fig.~\ref{fig:HExs}. Obviously, part of the pole shift 
to higher energies with respect to its nominal energy might be attributed to interference 
with the non-resonant background. Though, since the $N^{*}(1650)$ does not couple strongly 
to the $K\Lambda$ channel \cite{Tanabashi:2018oca} the $\pi N \rightarrow K\Lambda$ cross section data do not allow 
for a reliable location of the resonance pole position. The difference in reproducing appropriately 
the decay width is of less concern here for the same reason as in the $N^{*}(1535)$ case. 

Let us further have a look at couplings of the involved channels to the generated resonant 
states. Here we follow Ref.~\cite{Taylor:1972aa} and express the transition amplitude 
in the vicinity of the complex pole energy $z_{R} = E_R - {\rm i}\Gamma_R /2$ as
\beq
f_{jb,ia}(z) = f^{\rm BG}_{jb,ia}(z) -\frac{1}{2(q_{jb}q_{ia})^{1/2}} \frac{\beta_{jb} \beta_{ia}}{z-z_{R}} \;\;\; ,
\eeq{eq:Fres2}
where the non-resonant background contribution $f^{\rm BG}$ and the dependence of the resonant 
part on the on-shell c.m.~momenta $q_{jb}$ are shown explicitly. The complex couplings $\beta_{jb}$ 
can be determined from the residua of elastic 
scattering amplitudes calculated at the pole energy. They are related to the partial widths 
\beq
\Gamma_{jb} = \:\mid \beta_{jb} \mid ^{2}\: = \lim_{z \to z_R} \mid 2q_{jb}(z-z_R)f_{jb,jb}(z) \mid
\eeq{eq:G_i}
that refer to the decay into the $jb$ channel. 
The calculated partial decay widths are presented in Table \ref{tab:BR}, naturally 
only for channels that are open at the resonance energy. There, we note that in particular 
the decays of the $N^{*}(1535)$ resonance into the $\pi N$ channel are underestimated 
by our models A and B. There, the admixture of the $\eta_0 N$ component in the $\eta N$ state  
makes the relative disproportion between the $\pi N$ and $\eta N$ decay rates even bigger than the one 
observed for the C model. The C model does surprisingly well concerning the $\pi N$ partial width 
but the calculated $\eta N$ partial width seems to be too large. On the other hand, the models A and B 
provide quite reasonable decay rates of the $N^{*}(1650)$ state in a qualitative agreement 
with those reported by the PDG. There, only the $K\Lambda$ decay width appears to be a bit low 
but in accordance with a small coupling of the $K\Lambda$ channel to the $N^{*}(1650)$ resonance.

\begin{table}[htb]
\caption{Calculated partial decay widths $\Gamma_{jb}$ (in MeV) for the poles $z_1$ and $z_2$ related to the $N^{\star}(1535)$ 
and $N^{\star}(1650)$ resonances, respectively. The last line shows the decay widths estimated 
by the PDG \cite{Tanabashi:2018oca}.}
\label{tab:BR}     
\begin{center}
\begin{tabular}{c|ccc|ccc}
\hline\noalign{\smallskip}
         & \multicolumn{3}{c}{$z_1$ pole}  & \multicolumn{3}{c}{$z_2$ pole} \\
  model  & $\pi N$ & $\eta N$ & $K\Lambda$ & $\pi N$ & $\eta N$ & $K\Lambda$ \\
\noalign{\smallskip}\hline\noalign{\smallskip}
    A    & 14.4 & 42.2 &  --- &  87.2 & 35.7 & 3.11  \\
    B    & 16.1 & 38.3 &  --- &  90.7 & 40.4 & 3.45  \\
    C    & 48.2 & 87.0 &  --- &  ---  &  --- &  ---  \\
\noalign{\smallskip}\hline
PDG \cite{Tanabashi:2018oca} & 54.6 & 54.6 &  --- & 81.0 & 33.8 & 13.5 \\
\noalign{\smallskip}\hline
\end{tabular}
\end{center}
\end{table}

Additional insight on the formation of dynamically generated resonant states can be obtained 
from comparing how strongly the pertinent resonant poles couple to the involved channels including 
those that open at higher energies. For this purpose we define dimensionless couplings 
$\tilde{\beta}_{jb}$ as
\beq
\tilde{\beta}_{jb} = \beta_{jb} / (2q_{jb})^{1/2}  
\eeq{eq:beta}
that also relate directly to the residue of the elastic amplitude, Eq.~(\ref{eq:Fres2}), 
since ${\rm Res}_{z=z_R}f_{jb,jb}(z)=-\tilde{\beta_{jb}}^{\!2}$. The moduli of the $\tilde{\beta}_{jb}$ 
couplings are shown in Table~\ref{tab:beta}.

\begin{table}[htb]
\caption{Calculated moduli of the channel couplings $\mid \tilde{\beta}_j \mid$ for the poles 
$z_1$ and $z_2$ related to the $N^{\star}(1535)$ and $N^{\star}(1650)$ resonances, respectively.}
\label{tab:beta}     
\begin{center}
\begin{tabular}{c|ccccc|ccccc}
\hline\noalign{\smallskip}
         & \multicolumn{5}{c}{$z_1$ pole}              
         & \multicolumn{5}{c}{$z_2$ pole} \\
  model  & $\pi N$ & $\eta N$ & $K\Lambda$ & $K\Sigma$ & $\eta'N$ 
         & $\pi N$ & $\eta N$ & $K\Lambda$ & $K\Sigma$ & $\eta'N$ \\
\noalign{\smallskip}\hline\noalign{\smallskip}
    A    & 0.13 & 0.39 & 0.88 & 0.41 & 0.05 & 0.27 & 0.21 & 0.08 & 0.84 & 0.38 \\
    B    & 0.14 & 0.37 & 0.82 & 0.37 & 0.05 & 0.28 & 0.22 & 0.08 & 0.85 & 0.39 \\
    C    & 0.24 & 0.47 & 1.46 & 3.15 & 0.00 &  --- &  --- &  --- &  --- &  --- \\
\noalign{\smallskip}\hline
\end{tabular}
\end{center}
\end{table}

The $\tilde{\beta}_{\pi N}$ and $\tilde{\beta}_{\eta N}$ couplings appear to be reasonably stable, 
professing only marginal variation for a different treatment of the inelasticity factor 
$\epsilon_r(\sqrt{s})$ in models A and B. The $K\Lambda$ channel couples quite strongly 
to the $N^{*}(1535)$ related $z_1$ pole and rather weakly to the $N^{*}(1650)$ related $z_2$ pole. 
There is no pole that we could assign to the $N^{*}(1650)$ resonance within our C model and 
the state couples most strongly to the $K\Sigma$ channel for the A and B models. For the C model, 
it is also interesting to note a quite large coupling of the $K\Sigma$ channel to the $z_1$ pole 
dwarfing even the already large coupling of the $K\Lambda$ channel to the same pole. In general,
all four channels included in the {\it low energy} C model fit couple more strongly to the $z_1$ pole 
when compared with the couplings found for the models fitted to the data that cover the higher 
energies as well.

Finally, we have looked at the origin of the poles in a hypothetical situation when all inter-channel 
couplings are switched off, the so called zero coupling limit (ZCL) \cite{Eden:1964zz, Pearce:1988rk}.
Depending on the strength of the interaction in the decoupled channel, the ZCL pole 
can appear either as a bound state (on the physical RS) or as a resonance or a virtual state 
(on the unphysical RS). When the inter-channel interaction is gradually switched on, 
the pole moves along a continuous trajectory from its position in the ZCL to the position 
where we find it in the physical limit, with all inter-channel couplings at their physical 
values. The necessary conditions required for emergence of poles in the ZCL were discussed 
in some detail in \cite{Cieply:2016jby} where the pole movements were demonstrated for several 
chiral approaches to the $\bar{K}N$ coupled channels system. The analyticity of the $S$-matrix 
with respect to continuous variations of the model parameters guarantees that each pole found 
in the physical limit can have its origin traced to the ZCL, to a pole persisting 
in a single decoupled channel. 

\begin{figure}[htb]    
\centering
\includegraphics[width=0.48\textwidth]{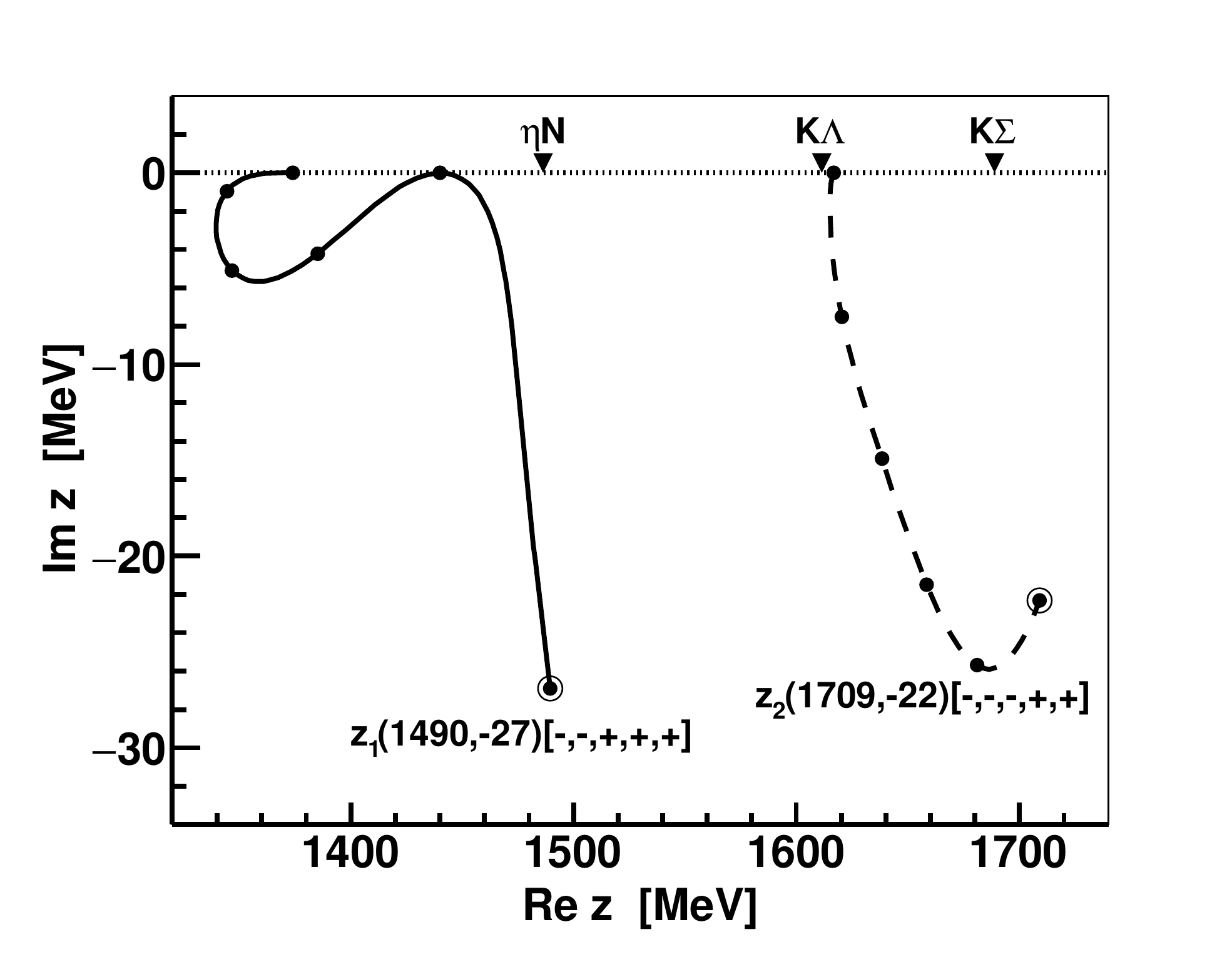}\hspace*{2mm}
\includegraphics[width=0.48\textwidth]{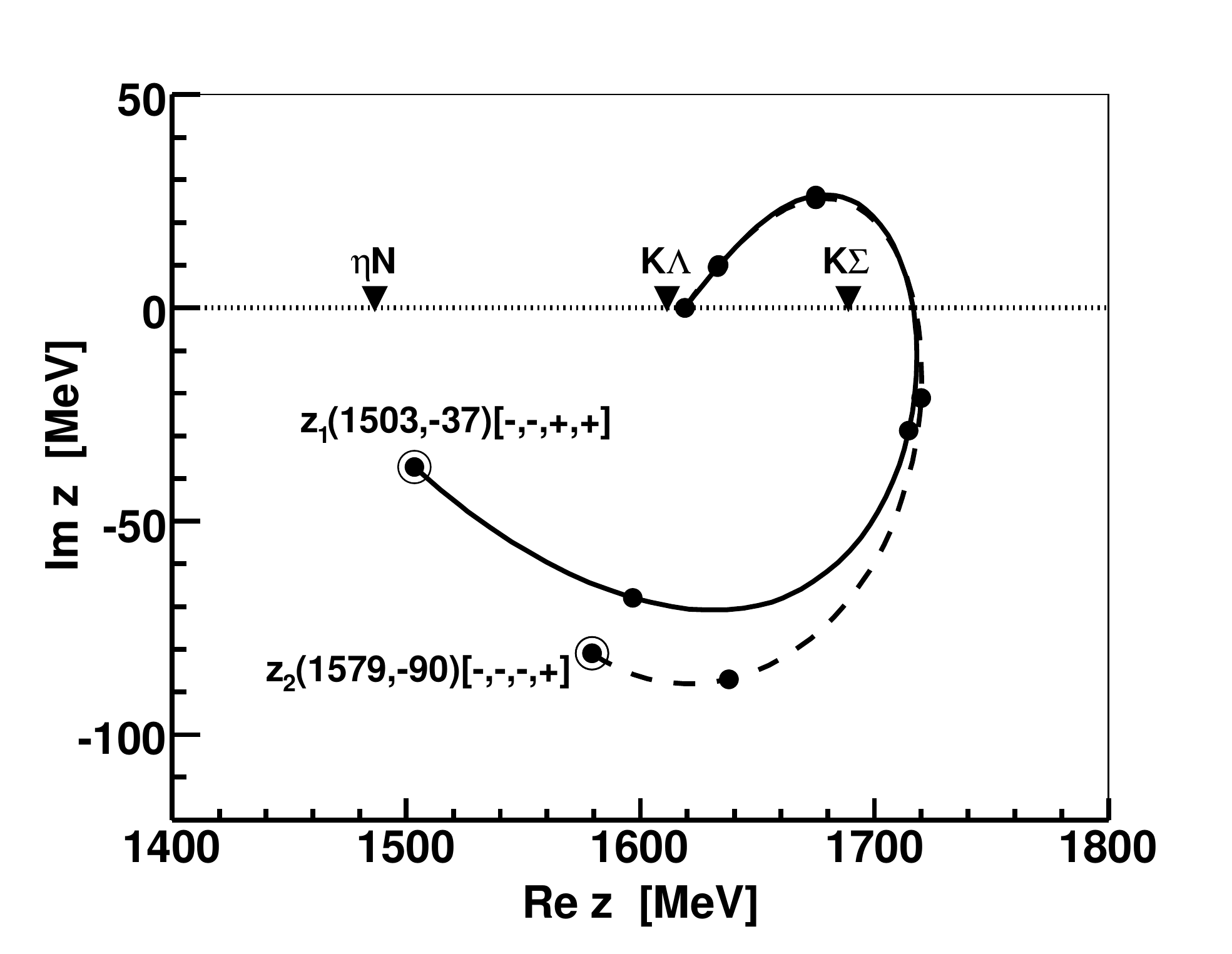} \\
\caption{Movement of the poles $z_1$ and $z_2$ upon gradually switching off the inter-channel couplings. 
The positions of the poles in a physical limit are encircled and marked by the labels that also denote 
the Riemann sheets the poles are located on. The small dots mark positions of the poles for the scaling 
factors from $x=0$ (zero coupling limit) to $x=1$ (physical limit) in steps of $0.2$. The triangles at 
the real axis point to the channel thresholds. Left panel: model A, right panel: CS model \cite{Cieply:2013sya}.}
\label{fig:ZCL}
\end{figure}

For the $\eta N$ coupled channels system the movement of the poles assigned to the $N^{*}(1535)$ 
and $N^{*}(1650)$ resonances was already looked at in \cite{Cieply:2013sya}. For a convenience 
and a direct comparison with our present findings we show the figure taken from Ref.~\cite{Cieply:2013sya} 
in the right panel of our Fig.~\ref{fig:ZCL}, aside an analogous analysis performed with
our model A that also contains poles assigned to both resonances. In both panels, the movement 
of the $z_1$ and $z_2$ poles is followed to their positions in the ZCL. This is visualised in Fig.~\ref{fig:ZCL}
on the Riemann sheets [-,-,+,+,+] (continuous lines, $z_1$ pole) and [-,-,-,+,+] 
(dashed lines, $z_2$ pole) in the lower half of the complex energy plane. 
The pole trajectories show the pole positions as we gradually decrease a scaling factor $x$ 
that is applied to the non-diagonal inter-channel couplings $C_{jb,ia}$ from $x=1$ 
(physical limit) to $x=0$ (ZCL). The dots mark the positions of the poles 
for $x=1$, $x=0.8$, ..., $x=0$ with the last point showing the final ZCL pole positions.
The initial pole positions in the physical limit ($x=1$ providing full physical couplings) 
are encircled and match those given in Table \ref{tab:poles} (for the left panel) 
or in \cite{Cieply:2013sya} (right panel).

First, let us have a look at the right panel. There, both poles originate from the same point, 
a virtual $K\Sigma$ state found in the ZCL at an energy about 70 MeV below the $K\Sigma$ threshold. 
It should be noted that when the pole trajectory passes across the real axis the pole continues its 
path on a RS that has reversed signs for all channels the branch cuts of were crossed, i.e. those 
with thresholds below the crossing point. For this reason the $z_1$ pole moves on the [+,+,-,-] RS
\footnote{Only four coupled channels were considered in \cite{Cieply:2013sya}, so the sign of the 
{\it decoupled} $\eta'N$ channel is omitted here.}
in the upper half of the figure (for $\mathrm{Im}\, z > 0$) and the $z_2$ pole on the [+,+,+,-] RS, 
both of them reaching the ZCL at the unphysical sheet in the $K\Sigma$ channel. 
Since the $z_1$ and $z_2$ poles emerge from the same point in the ZCL, they are shadow 
poles one to each other. In fact, there are even more shadow poles that evolve on different 
Riemann sheets (RSs) from the same ZCL position. As soon as the inter-channel couplings switch 
on the ZCL pole departs from the real axis and can start moving on any of the RSs
that keep the minus sign for the $K\Sigma$ channel. For small values of the scaling factor $x$ 
the pole positions on these RSs remain relatively close. However, the inter-channel dynamics 
may lead to large differences between the positions of the shadow poles in the physical limit 
(or for any large $x$). The physics at the real axis is always affected most strongly 
by the nearest of the shadow poles. In principle, it may even be a pole 
on a more distant RS than the second one which is commonly looked at. We have checked 
that the $z_1$ and $z_2$ poles reported here are the closest ones relevant for the energies 
in the region of the $N^{*}(1535)$ and $N^{*}(1650)$ resonances.

In the left panel of Fig.~\ref{fig:ZCL} we see that for model A the poles assigned 
to $N^{*}(1535)$ and $N^{*}(1650)$ originate from different positions in the ZCL and 
the trajectories of the $z_1$ and $z_2$ poles do not cross the real axis, staying  
in the [-,-,+,+,+] and [-,-,-,+,+] RSs, respectively. Both features make the A model 
different from the CS model reported in \cite{Cieply:2013sya}. While the origin 
of the A model $z_2$ pole can be traced to the $K\Sigma$ bound state in the ZCL 
the path of the $z_1$ pole movement (starting from its physical $N^{*}(1535)$ position) 
goes very fast to the real axis, then moves below it to reach the ZCL as an $K\Lambda$ 
bound state. The bound state found in the decoupled $K\Lambda$ channel is enabled 
by a relatively large and attractive diagonal coupling $C_{K\Lambda, K\Lambda}$ 
in our model A as well as by a large coupling of the $K\Lambda$ channel to the $z_1$ pole 
seen in Table~\ref{tab:beta}. In fact, we found that for the A model, all diagonal couplings 
$C_{jb,jb}$, with the exception of the $\pi N$ one, are large and attractive. 
This reflects relatively large NLO contributions emerging from the Lagrangian terms 
proportional to the $d$-couplings.

As the $\eta'N$ channel was not considered in \cite{Cieply:2013sya} it seems natural 
to relate the qualitative differences observed for the pole movements in the left 
and right panels of Fig.~\ref{fig:ZCL} to the inclusion of the $\eta'N$ channel 
in our approach, i.e.~to the $\eta_0$ admixture in the $\eta N$ interaction. However,
our new model C provides for another solution (local minimum in the $\chi^{2}$ fits) 
that behaves differently even from the CS model, not only missing completely 
on the $N^{*}(1650)$ pole. We have found that the $z_1$ pole trajectory does differ 
significantly from the one observed in right panel of Fig.~\ref{fig:ZCL}, the pole drifting 
fast to quite large energies and far from the real axis, making it difficult to follow it 
to the ZCL. On the other hand, we were able to reproduce qualitatively the CS model results 
(including the $z_2$ pole and its movement to ZCL) when we fitted the same set of experimental 
data as in \cite{Cieply:2013sya}. Thus, we conclude that the characteristics of the poles 
assigned to $N^{*}(1535)$ and $N^{*}(1650)$ are not so reliably established when the fitted 
experimental data are limited to low energies.

We close this section with a comment on the $\eta'N$ interaction. When analyzing the poles 
in the ZCL we found that the A model diagonal coupling for the $\eta'N$ channel 
is also sufficient to generate a bound state in the decoupled $\eta'N$ channel. 
However, when the inter-channel couplings are switched 
on the pole moves quickly to high energies (above 2 GeV) and far away from the real axis, 
so it does not have any impact on physical observables, at least not at energies covered 
in the current work. In principle, it might be possible to tune the model parameters to keep 
the pole close to the $\eta'N$ threshold even in the physical limit and assign it 
to the $N^{*}(1895)$ resonance. If this was achieved the $\eta'N$ interaction would 
become attractive, contrary to our current predictions and in line with the indirect 
evidence discussed in the previous section in relation to the elastic $\eta'N$ amplitude 
presented in Fig.~\ref{fig:AetapN}. The option of generating a $N^{*}(1895)$ resonance 
dynamically within our approach cannot be ruled out, the possibility is evidently there. 
However, we have not managed to keep the $\eta'N$ pole in a physically relevant region 
by simple modifications of the A model, playing either with the diagonal $C_{\eta'N,\eta'N}$ 
coupling or with the $\alpha_{\eta'N}$ inverse range.

\section{Summary}
\label{sec:sum}

We have presented a coupled-channels model that describes the s-wave interactions 
of pseudoscalar mesons with the lightest baryons in the strangeness $S=0$ sector 
and includes the $\eta_0 - \eta_8$ mixing. The inter-channel couplings were derived 
from the chiral Lagrangian formulated up to the ${\cal O}(p^2)$ order and with 
its model parameters (LECs and inverse ranges) fitted to the available $\pi N$ amplitudes 
and to low energy cross section data covering quite broad interval of energies 
up to about 2 GeV. The approach utilizes the Yamaguchi form factors to regularize 
the intermediate state loop functions and provides a natural extension off the energy shell 
making the resulting separable meson-baryon amplitudes suitable for in-medium applications. 
The Lippmann-Schwinger equation was used to sum the major part of the ChPT perturbation 
series and to guarantee unitarity of the scattering $T$-matrix. Despite a relative simplicity 
of the model and its restriction to a selected set of two-particle coupled channels 
it provides a satisfactory description of the low-energy experimental data as well as 
some interesting predictions for the $\eta N$ and $\eta'N$ systems 
and for the related $N^{*}(J^{P}=1/2^{-})$ resonant states.

An explicit inclusion of the singlet meson field $\eta_0$ leads to more attractive $\eta N$ interaction 
at energies close to the channel threshold, a feature quite relevant for theoretical predictions and 
possible observation of the $\eta$-nuclear bound states \cite{Cieply:2013sga}. As far as we know we are 
the first to demonstrate this behaviour, though it was already foreseen in \cite{Bass:2005hn}. 
The real part of the $\eta N$ scattering length predicted by our A model, $\mathrm{Re}\, a_{\eta N} = 1.05$ fm, 
is significantly larger than the values obtained without the inclusion of the $\eta'N$ channel, 
either the $\mathrm{Re}\, a_{\eta N} = 0.73$ fm prediction by our C model or quite similar values reported 
earlier in \cite{Cieply:2013sya} and \cite{Nieves:2001wt}. We also note that the large model A value 
is compatible with the phenomenological $K$-matrix evalulation of the $\eta N$ scattering 
length by Green and Wycech \cite{Green:2004tj}.

The $N^{*}(1535)$ and $N^{*}(1650)$ resonances are generated dynamically within our coupled-channel 
approach with strong couplings to the $K\Lambda$ and $K\Sigma$ channels, respectively. When the $\eta'N$ 
channel is decoupled (or the $\eta_0$ singlet excluded) the pole assigned to the $N^{*}(1650)$ may be missing 
as demonstrated by our C model, though we were also able to reproduce the results of \cite{Cieply:2013sya}
where both poles were present and originate from the same $K\Sigma$ virtual state. On the other hand, 
our model A results show that the inclusion of the $\eta_0$ field leads to large diagonal couplings 
in the $K\Lambda$ and $\eta'N$ 
channels, sufficient to generate bound states in the ZCL. For the inter-channel couplings restored 
to their physical values the $K\Lambda$ pole can be identified with the $N^{*}(1535)$ state while 
the $\eta'N$ pole drifts far away from the real axis and to energies beyond 2 GeV making it irrelevant, 
at least within our model A setting. We still find it intriguing that such an $\eta'N$ ZCL pole is there 
and might be related to the debated $N^{*}(1895)$ resonance provided a suitable parameter set was found to keep 
the pole in the physically relevant region even when the inter-channel couplings are switched on.

Finally, our models predict a repulsive $\eta'N$ interaction in a broad interval of energies 
around the channel threshold. Although the $\eta'N$ scattering length predicted by all our models, 
with $\mathrm{Re}\, a_{\eta'N} = -0.4$~fm, falls within the limits derived from the $pp \longrightarrow pp\eta'$ 
experiment at COSY \cite{Czerwinski:2014yot}, the repulsive character of the interaction 
is at odds with indications based on the in-medium $\eta'$ mass shift observed in photoproduction 
experiments on nuclear targets \cite{Nanova:2016cyn}. However, due to in-built limitations 
of our approach and non-sufficient experimental input at the relevant energies our predictions 
for the $\eta'N$ amplitude may not be conclusive. In particular, one should seriously consider
adding other channels such as the $\pi\pi N$ one, vector-baryon channels considered 
in e.g.~\cite{Oset:2010ub}, or couplings to some relevant resonant states not generated 
dynamically within the present approach.

\section*{Acknowledgement}

We thank J.~Mare\v{s} for careful reading of the manuscipt, encouragement and comments. 
This work was supported by the Czech Science Foundation GACR grant 19-19640S.

\appendix

\setcounter{figure}{0}

\section{Chiral building blocks, notation and conventions}
\label{app:chptnom}

The Goldstone bosons octet field $\phi_{8}$ and flavor-singlet meson field $\phi_{0}$ 
are collected in a matrix $U=\exp\left({\rm i}\sqrt{2}\Phi/F_{0}\right)$, 
where $\Phi = \phi_{8} + \phi_{0}$,  $F_{0}$ stands for the meson decay constant 
in the SU(3) chiral limit of vanishing light-quark masses, $m_{u,d,s}\rightarrow 0$, 
$F_{0}\approx 80\,\mathrm{MeV}$ \cite{Aoki:2016frl}, and
\beqa
\phi_{0} &=& (\eta_{0}/\sqrt{3})\,\mathds{1}_{3\times 3} \; , \CR[\MySep]
\phi_{8} &=& \left(
\begin{array}{ccc}
  \frac{1}{\sqrt{2}}\pi^{0}+\frac{1}{\sqrt{6}}\eta_{8} & \pi^{+} & K^{+} \\
  \pi^{-} & -\frac{1}{\sqrt{2}}\pi^{0}+\frac{1}{\sqrt{6}}\eta_{8} & K^{0} \\
  K^{-} & \bar K^{0} & -\frac{2}{\sqrt{6}}\eta_{8} 
\end{array}
\right) \; .
\eeqa{eq:Phi}
We also define $u=\sqrt{U}$, and
\begin{eqnarray*}
\nabla_{\mu}U &=& \partial_{\mu}U-{\rm i}(v_{\mu}+a_{\mu})U+{\rm i}U(v_{\mu}-a_{\mu})   \,,\quad 
                  u_{\mu}={\rm i}u^{\dagger}\left(\nabla_{\mu}U\right)u^{\dagger} \,,\\
\Gamma^{\mu} &=& \frac{1}{2}\left(u^{\dagger}[\partial^{\mu}-{\rm i}(v^{\mu}+a^{\mu})]u + 
                  u[\partial^{\mu}-{\rm i}(v^{\mu}-a^{\mu})]u^{\dagger}\right)\,, \\
\chi      &=& 2B_{0}(s+{\rm i}p) \,,\quad 
\chi_{\pm} = \left(u^{\dagger}\chi u^{\dagger}\pm u\chi^{\dagger}u\right)
\end{eqnarray*}
where $v$, $a$, $s$ (not to be confused with the Mandelstam $s$), and $p$ denote the vector, 
axial-vector, scalar and pseudoscalar source fields, respectively, and $B_{0}$ stands 
for a low-energy constant related to the light-quark condensate in the chiral limit 
\cite{Gasser:1984gg,Aoki:2016frl}. In this work, we set $s=\mathrm{diag}(\hat{m},\hat{m},m_{s})$, $p=v=a=0$, 
where $\hat{m}=\frac{1}{2}(m_{u}+m_{d})$ is taken as an average of the {\it up} and {\it down} 
quark masses. 

Concerning the notation used in Eqs.~(\ref{eq:M_LagrBMW}-\ref{eq:MB_LagrBMW}) we also mention that 
the brackets $\langle\cdots\rangle$ represent the trace in flavor space and the baryon-octet mass 
in the three-flavor chiral limit is denoted by $\overset{\circ}{m}$. 

The following expansions in the meson-matrix field $\Phi$ can be useful:
\begin{eqnarray}
u_{\mu} &=& 2a_{\mu}-\frac{\sqrt{2}}{F_{0}}\left(\partial_{\mu}\Phi +{\rm i}\lbrack \Phi,\,v_{\mu}\rbrack\right) 
            -\frac{1}{2F_{0}^2}\lbrack\Phi,\,\lbrack\Phi,\,a_{\mu}\rbrack\rbrack \nonumber \\ 
        &+& \frac{\sqrt{2}}{12F_{0}^3}\lbrack\Phi,\,\lbrack\Phi,\,\left(\partial_{\mu}\Phi 
            +{\rm i}\lbrack \Phi,\,v_{\mu}\rbrack\right)\rbrack\rbrack + \ldots\,,\label{eq:umuExpanded} \\
\Gamma^{\mu} &=& -{\rm i}v^{\mu} - \frac{1}{\sqrt{2}F_{0}}\lbrack\Phi,\,a^{\mu}\rbrack 
                 +\frac{1}{4F_{0}^{2}}\lbrack\Phi,\,\left(\partial^{\mu}\Phi +{\rm i}\lbrack \Phi,\,v^{\mu}\rbrack\right)\rbrack + \ldots 
                 \,,\label{eq:GammaExpanded} \\
\chi_{+} &=& 4B_{0}s + \frac{2\sqrt{2}B_{0}}{F_{0}}\lbrace\Phi,\,p\rbrace 
             - \frac{B_{0}}{F_{0}^{2}}\lbrace\Phi,\,\lbrace\Phi,\,s\rbrace\rbrace 
             - \frac{\sqrt{2}B_{0}}{6F_{0}^{3}}\lbrace\Phi,\,\lbrace\Phi,\,\lbrace\Phi,\,p\rbrace\rbrace\rbrace  \nonumber \\ 
         &+& \frac{B_{0}}{24F_{0}^{4}}\lbrace\Phi,\,\lbrace\Phi,\,\lbrace\Phi,\,\lbrace\Phi,\,s\rbrace\rbrace\rbrace\rbrace 
             + \ldots\,,\label{eq:chiplusExpanded} \\
\chi_{-} &=& 4{\rm i}B_{0}p -\frac{2\sqrt{2}{\rm i}B_{0}}{F_{0}}\lbrace\Phi,\,s\rbrace 
             - \frac{{\rm i}B_{0}}{F_{0}^2}\lbrace\Phi,\,\lbrace\Phi,\,p\rbrace\rbrace 
             + \ldots\,.\label{eq:chiminusExpanded}
\end{eqnarray}

Finally, the baryon fields are collected in the matrix
\beq
B = \left(
  \begin{array}{ccc}
  \frac{1}{\sqrt{2}}\Sigma^{0}+\frac{1}{\sqrt{6}}\Lambda & \Sigma^{+} & p \\
  \Sigma^{-} & -\frac{1}{\sqrt{2}}\Sigma^{0}+\frac{1}{\sqrt{6}}\Lambda & n \\
  \Xi^{-} & \Xi^{0} & -\frac{2}{\sqrt{6}}\Lambda \end{array}
    \right) \;,
\eeq{eq:B}
and the covariant derivative $D^{\mu}$ acts as $\lbrack D^{\mu},B\rbrack:=\partial^{\mu}B + \lbrack\Gamma^{\mu},B\rbrack$.

\section{Isospin decomposition and channel matrix notation}
\label{app:channels}

Let us first consider meson-baryon scattering in the $I=1/2$, $S=0$ sector. The channels $|jb\rangle$ 
are ordered according to their threshold energies as 
\begin{displaymath}
|\pi N\rangle,\; |\eta N\rangle,\; |K\Lambda\rangle,\; |K\Sigma\rangle,\; |\eta'N\rangle \; .
\end{displaymath}
For the isospin states $|I,I_{3}\rangle$, we use the convention where there are minus signs 
in the states $|\pi^{+}\rangle = -|1,1\rangle_{\pi}$, $|\bar K^{0}\rangle = -|\frac{1}{2},\frac{1}{2}\rangle_{\bar K}$, 
$|\Sigma^{+}\rangle = -|1,1\rangle_{\Sigma}$ and $|\Xi^{0}\rangle = -|\frac{1}{2},\frac{1}{2}\rangle_{\Xi}$, 
which is consistent with the parameterizations of the corresponding field operators in Eqs.~(\ref{eq:Phi}), 
(\ref{eq:B}) and the usual phase conventions for the Clebsch-Gordan coefficients. We then find e.g.
\begin{eqnarray*}
 \left|\frac{1}{2},+\frac{1}{2}\right>_{\pi N} \,&=&\, -\left(\sqrt{\frac{2}{3}}|\pi^{+}n\rangle + \sqrt{\frac{1}{3}}|\pi^{0}p\rangle\right)\,,\\
 \left|\frac{1}{2},-\frac{1}{2}\right>_{\pi N} \,&=&\, \sqrt{\frac{1}{3}}|\pi^{0}n\rangle - \sqrt{\frac{2}{3}}|\pi^{-}p\rangle\,,\\
 \left|\frac{1}{2},+\frac{1}{2}\right>_{K\Sigma} \,&=&\, \sqrt{\frac{2}{3}}|K^{0}\Sigma^{+}\rangle + \sqrt{\frac{1}{3}}|K^{+}\Sigma^{0}\rangle\,,\\
 \left|\frac{1}{2},-\frac{1}{2}\right>_{K\Sigma} \,&=&\, \sqrt{\frac{2}{3}}|K^{+}\Sigma^{-}\rangle - \sqrt{\frac{1}{3}}|K^{0}\Sigma^{0}\rangle\,.   
\end{eqnarray*}

The $I=3/2$ sector consists of only two channels,
\begin{displaymath}
|\pi N\rangle,\; |K\Sigma\rangle \; ,
\end{displaymath}
and it is simplest to compute the amplitudes for 
\begin{eqnarray*}
\left|\frac{3}{2},-\frac{3}{2}\right>_{\pi N}   &=& \left|\pi^{-}n\right> \; ,\\ 
\left|\frac{3}{2},-\frac{3}{2}\right>_{K\Sigma} &=& \left|K^{0}\Sigma^{-}\right> \; .
\end{eqnarray*}

Throughout the paper we often employ a matrix formalism with the matrix indices $jb$ 
running over the coupled channels space, five channels for $I=1/2$ and two channels in the $I=3/2$ sector. 
The matrices comprise entries for the meson-baryon reactions $(ia)\rightarrow(jb)$,   
with $a,b$ and $i,j$ standing for the baryon and meson species, respectively.
In this matrix notation the baryon-mass matrix $m$ is diagonal with elements 
$m_{jb,ia}=\delta_{ba}\,\delta_{ji}\,m_{b}$. Explicitly, 
\begin{eqnarray*}
\begin{array}{ll}
m=\mathrm{diag}(m_{N},m_{N},m_{\Lambda},m_{\Sigma},m_{N}) & {\rm for}\; I=1/2 \\
m=\mathrm{diag}(m_{N},m_{\Sigma})                         & {\rm for}\; I=3/2 \; .
\end{array}
\end{eqnarray*}
In the same way, we introduce a meson mass matrix $M$, and a diagonal matrix $E$ containing the baryon center-of-mass energies,
\begin{equation}
E = \frac{s+m^2-M^2}{2\sqrt{s}} \; .
\end{equation}
Similarly, when appropriate, the Mandelstam variable $s$ is also understood to acquire the matrix 
form $s\,\mathds{1}$, with $\mathds{1}$ denoting the unit matrix in the channel space. 
Finally, a diagonal matrix $F_{\Phi}\approx F_{0}\,\mathds{1}$ is introduced collecting 
the meson decay constants corresponding to our meson-baryon channels,
\begin{eqnarray*}
\begin{array}{ll}
F_{\Phi} = \mathrm{diag}(F_{\pi},F_{\eta},F_{K},F_{K},F_{\eta'}) & {\rm for}\; I=1/2 \\
F_{\Phi} = \mathrm{diag}(F_{\pi},F_{K})                          & {\rm for}\; I=3/2 \; .
\end{array}
\end{eqnarray*}
The meaning of inverses and square roots of these diagonal matrices is self-evident.

\section{Channel matrices}
\label{app:chanmats}

\subsection{The isospin $I=1/2$ sector}

For the Weinberg-Tomozawa (WT) interaction term derived from the chiral connection 
in the Lagrangian (\ref{eq:MB_LagrBMW}), one finds
\begin{equation}
C_{WT} = \Delta_{\vartheta}^{8} 
\left( \begin{array}{ccccc} 
       2       &      0       &  \frac{3}{2} & -\frac{1}{2} &      0       \\[\MySep] 
       0       &      0       & -\frac{3}{2} & -\frac{3}{2} &      0       \\[\MySep]
   \frac{3}{2} & -\frac{3}{2} &      0       &      0       & -\frac{3}{2} \\[\MySep] 
  -\frac{1}{2} & -\frac{3}{2} &      0       &      2       & -\frac{3}{2} \\[\MySep] 
       0       &      0       & -\frac{3}{2} & -\frac{3}{2} &      0 
\end{array} \right) 
\Delta_{\vartheta}^{8} \;,
\end{equation}
for the coupling matrix appearing in Eq.~(\ref{eq:f0pPOT2}), where the diagonal matrices
\begin{equation}
\Delta_{\vartheta}^{8} = \mathrm{diag}(1,\cos\vartheta,1,1,\sin\vartheta)\,, \quad \Delta_{\vartheta}^{0}=\mathrm{diag}(1,-\sin\vartheta,1,1,\cos\vartheta)
\end{equation}
are introduced to account for the singlet-octet $\eta$ mixing parameterized 
by the mixing angle $\vartheta$. We note in passing that the entries of $1$
in $\Delta_{\vartheta}^{0}$ are actually irrelevant due to the particular form of the channel matrices, as given below.
The $w_s$ matrix reads as
\begin{equation}
C_{w_{s}} = \Delta_{\vartheta}^{0} \,\mathcal{M}_{\eta} \,\Delta_{\vartheta}^{0}\,,
\end{equation}
and to simplify some notation we also introduce an auxiliary matrix
\beq
\mathcal{M}_{\eta} = \left( \begin{array}{ccccc} 
  0 & 0 & 0 & 0 & 0 \\ 
  0 & 1 & 0 & 0 & 1 \\ 
  0 & 0 & 0 & 0 & 0 \\ 
  0 & 0 & 0 & 0 & 0 \\ 
  0 & 1 & 0 & 0 & 1 
\end{array} \right) \; .
\eeq{}
The various coupling matrices are specified below anticipating that
\begin{eqnarray*}
C_{\pi} &=& C_{\pi,b}+C_{\pi,c} \; ,\quad C_{K} = C_{K,b}+C_{K,c} \; ,\quad C_{d} = C_{d,14}+C_{d,57} \; , \\ 
C_{\cdot,\cdot} &=& \Delta_{\vartheta}^{8}C_{\cdot,\cdot}^{88}\Delta_{\vartheta}^{8} 
                 +  \Delta_{\vartheta}^{0}C_{\cdot,\cdot}^{08}\Delta_{\vartheta}^{8} 
                 +  \Delta_{\vartheta}^{8}C_{\cdot,\cdot}^{80}\Delta_{\vartheta}^{0} 
                 +  \Delta_{\vartheta}^{0}C_{\cdot,\cdot}^{00}\Delta_{\vartheta}^{0} \,,
\end{eqnarray*}
where the dots stand for the coupling matrix indices $\pi$, $K$, $d$, $s$ and $u$ or for their parts in the splitting given in
the first line of equations above.

The components of the $C_{\pi}$, $C_{K}$ and $C_{d}$ matrices read as follows:

\begin{eqnarray*}
C_{\pi,b}^{88} \!\!\! &=& \!\! \left( \!
\begin{array}{ccccc} 
   2\BeNul \!+\! \BeDe \!+\! \BeFe  &\!        
  -(\BeDe \!+\! \BeFe)              &\!
   \frac{1}{4}(\BeDe \!+\! 3\BeFe)  &\!
   \frac{1}{4}(\BeDe \!-\! \BeFe)   &\!
  -(\BeDe \!+\! \BeFe)              \\[\MySep]
  \ldots                                          &\!
  -\frac{1}{3}(2\BeNul \!+\! 3\BeDe \!-\! 5\BeFe) &\!
  -\frac{1}{4}(\BeDe \!+\! 3\BeFe)                &\!
   \frac{3}{4}(\BeDe \!-\!  \BeFe)                &\!
  -\frac{1}{3}(2\BeNul \!+\! 3\BeDe \!-\! 5\BeFe) \\[\MySep]
  \ldots                              &\!
  \ldots                              &\!
   0                                  &\!
   0                                  &\!
  -\frac{1}{4}(\BeDe \!+\! 3\BeFe)    \\[\MySep]
  \ldots                              &\!
  \ldots                              &\!
  \ldots                              &\!
   0                                  &\!
   \frac{3}{4}(\BeDe \!-\! \BeFe)     \\[\MySep]
  \ldots                              &\!
  \ldots                              &\!
  \ldots                              &\!
  \ldots                              &\!
 -\frac{1}{3}(2\BeNul \!+\! 3\BeDe \!-\! 5\BeFe)
\end{array}
\!\right) \; ,\\[\MySep]
C_{\pi,b}^{08} \!\!\! &=& \!\! -\sqrt{2}
\left(
\begin{array}{ccccc} 
  0 & 0 & 0 & 0 & 0 \\[\MySep] 
  b_{D}\!+\!b_{F} & -\frac{1}{3}(4b_{0}\!+\!3b_{D}\!-\!b_{F}) & 0 & 0 & -\frac{1}{3}(4b_{0}\!+\!3b_{D}\!-\!b_{F}) \\[\MySep]
  0 & 0 & 0 & 0 & 0 \\[\MySep] 
  0 & 0 & 0 & 0 & 0 \\[\MySep] 
  b_{D}\!+\!b_{F} & -\frac{1}{3}(4b_{0}\!+\!3b_{D}\!-\!b_{F}) & 0 & 0 & -\frac{1}{3}(4b_{0}\!+\!3b_{D}\!-\!b_{F})  
\end{array}
\right) \; ,  \\[\MySep]
C_{\pi,b}^{80} \!\!\! &=& \!\! (C_{\pi,b}^{08})^{T} \;,  \\[\MySep]
C_{\pi,b}^{00} \!\!\! &=& \!\! \frac{2}{3}(b_{0}+2b_{F})\, \mathcal{M}_{\eta}\,,
\end{eqnarray*}

\begin{eqnarray*}
C_{K,b}^{88} \!\!\! &=& \!\! \left( \!
\begin{array}{ccccc} 
  0 & \! 
  0 & \!
  \frac{1}{4}(b_{D}\!+\!3b_{F}) & \!
  \frac{1}{4}(b_{D}\!-\!b_{F}) & \!
  0 \\[\MySep]
  \ldots & \!
  \frac{8}{3}(b_{0}\!+\!b_{D}\!-\!b_{F}) & \!
  \frac{5}{12}(b_{D}\!+\!3b_{F}) & \!
  -\frac{5}{4}(b_{D}\!-\!b_{F}) & \!
  \frac{8}{3}(b_{0}\!+\!b_{D}\!-\!b_{F}) \\[\MySep]
  \ldots & \!
  \ldots & \!
  \frac{1}{3}(6b_{0}\!+\!5b_{D}) & \!
  b_{D} & \!
  \frac{5}{12}(b_{D}\!+\!3b_{F}) \\[\MySep]
  \ldots  & \!
  \ldots  & \!
  \ldots  & \!
  2b_{0}+b_{D}\!-\!2b_{F} & \!
  -\frac{5}{4}(b_{D}\!-\!b_{F}) \\[\MySep]
  \ldots & \!
  \ldots & \!
  \ldots & \!
  \ldots & \!
  \frac{8}{3}(b_{0}\!+\!b_{D}-b_{F}) 
\end{array}
\! \right) \;, \\[\MySep]
C_{K,b}^{08} \!\!\! &=& \!\! -\sqrt{2}\left( \!
\begin{array}{ccccc} 
  0 & 0 & 0 & 0 & 0 \\ 
  0 & \frac{4}{3}(b_{0}\!+\!b_{D}\!-\!b_{F}) & \frac{1}{3}(b_{D}\!+\!3b_{F}) & -(b_{D}\!-\!b_{F}) & \frac{4}{3}(b_{0}\!+\!b_{D}\!-\!b_{F}) \\ 
  0 & 0 & 0 & 0 & 0 \\ 
  0 & 0 & 0 & 0 & 0 \\ 
  0 & \frac{4}{3}(b_{0}\!+\!b_{D}\!-\!b_{F}) & \frac{1}{3}(b_{D}\!+\!3b_{F}) & -(b_{D}\!-\!b_{F}) & \frac{4}{3}(b_{0}\!+\!b_{D}\!-\!b_{F})  
\end{array}
\!\right) \;, 
\end{eqnarray*}
\begin{eqnarray*}
C_{K,b}^{80} \!\!\! &=& \!\! (C_{K,b}^{08})^{T} \;, \\[\MySep]
C_{K,b}^{00} \!\!\! &=& \!\! \frac{4}{3}(b_{0}+b_{D}-b_{F})\, \mathcal{M}_{\eta} \;, \\[5mm]
C_{\pi,c}^{88} \!\!\! &=& \!\! 0 \;, \\[\MySep]
C_{\pi,c}^{08} \!\!\! &=& \!\! \sqrt{3}\left(
\begin{array}{ccccc} 
  0 & 0 & 0 & 0 & 0 \\ 
  c_{D}\!+\!c_{F} & -\frac{1}{3}(4c_{0}\!+\!3c_{D}\!-\!c_{F}) & 0 & 0 & -\frac{1}{3}(4c_{0}\!+\!3c_{D}\!-\!c_{F}) \\ 
  0 & 0 & 0 & 0 & 0 \\ 
  0 & 0 & 0 & 0 & 0 \\ 
  c_{D}\!+\!c_{F} & -\frac{1}{3}(4c_{0}\!+\!3c_{D}\!-\!c_{F}) & 0 & 0 & -\frac{1}{3}(4c_{0}\!+\!3c_{D}\!-\!c_{F})  
\end{array}
\right) \;, \\[\MySep]
C_{\pi,c}^{80} \!\!\! &=& \!\! (C_{\pi,c}^{08})^{T} \;, \\[\MySep]
C_{\pi,c}^{00} \!\!\! &=& \!\! -2\sqrt{\frac{2}{3}}(c_{0}+2c_{F})\, \mathcal{M}_{\eta} \;,
\end{eqnarray*}

\begin{eqnarray*}
C_{K,c}^{88} &=& 0 \;, \\[\MySep] 
C_{K,c}^{08} &=& \sqrt{3}\left(
\begin{array}{ccccc} 
  0 & 0 & 0 & 0 & 0 \\ 
  0 & \frac{4}{3}(c_{0}\!+\!c_{D}\!-c_{F}) & \frac{1}{3}(c_{D}\!+\!3c_{F}) & -(c_{D}\!-\!c_{F}) & \frac{4}{3}(c_{0}\!+\!c_{D}\!-\!c_{F}) \\ 
  0 & 0 & 0 & 0 & 0 \\ 
  0 & 0 & 0 & 0 & 0 \\ 
  0 & \frac{4}{3}(c_{0}\!+\!c_{D}\!-\!c_{F}) & \frac{1}{3}(c_{D}\!+\!3c_{F}) & -(c_{D}\!-\!c_{F}) & \frac{4}{3}(c_{0}\!+\!c_{D}\!-\!c_{F})  
\end{array}
\right) \;, \\[\MySep] 
C_{K,c}^{80} &=& (C_{K,c}^{08})^{T} \;, \\[\MySep] 
C_{K,c}^{00} &=& -4\sqrt{\frac{2}{3}}(c_{0}+c_{D}-c_{F})\, \mathcal{M}_{\eta} \;,
\end{eqnarray*}

\begin{eqnarray*}
C_{d,14}^{88} \!\!\! &=& \!\!\! \left( \!\!
\begin{array}{ccccc} 
  d_{1}\!+\!d_{2}\!+\!2d_4 & -(d_{1}\!+\!3d_{2}) & \frac{3}{2}(d_{1}\!+\!d_{2}) & -\frac{1}{2}(d_{1}\!-\!7d_{2}\!+\!2d_{3}) & -(d_{1}\!+\!3d_{2}) \\
  \ldots & -d_{1}\!+\!3d_{2}\!+\!2d_{4} & \frac{1}{2}(d_{1}\!-\!3d_{2}\!+\!2d_{3}) & \frac{1}{2}(d_{1}\!-\!3d_{2}) & -d_{1}\!+\!3d_{2}\!+\!2d_{4} \\
  \ldots & \ldots & 3d_{2}+2d_{4} & 3d_{2} & \frac{1}{2}(d_{1}-3d_{2}+2d_{3}) \\
  \ldots & \ldots & \ldots & -2d_{1}\!+\!d_{2}\!+\!2d_{4} & \frac{1}{2}(d_{1}\!-\!3d_{2}) \\
  \ldots & \ldots & \ldots & \ldots & -d_{1}\!+\!3d_{2}\!+\!2d_{4} 
\end{array}
\!\!\right) \;, \\[\MySep]
C_{d,14}^{08} \!\!\! &=& \!\! -\sqrt{2}d_{1}\left(
\begin{array}{ccccc} 
  0 &  0 & 0 & 0 &  0 \\ 
  1 & -1 & 1 & 1 & -1 \\ 
  0 &  0 & 0 & 0 &  0 \\ 
  0 &  0 & 0 & 0 &  0 \\ 
  1 & -1 & 1 & 1 & -1  
\end{array}\right) \;, \\[\MySep] 
C_{d,14}^{80} \!\!\! &=& \!\! (C_{d,14}^{08})^{T} \;, \\[\MySep] 
C_{d,14}^{00} \!\!\! &=& \!\! 2d_{4}\, \mathcal{M}_{\eta} \,,
\end{eqnarray*}

\begin{eqnarray*}
C_{d,57}^{08} \!&=&\! -\sqrt{2}\left(
\begin{array}{ccccc} 
  0 & 0 & 0 & 0 & 0 \\ 
  \frac{3}{2}(d_{5}\!+\!d_{6}) & \frac{1}{2}(d_{5}\!-\!3d_{6}) & \frac{1}{2}(d_{5}\!+\!3d_{6}) & 
  -\frac{3}{2}(d_{5}\!-\!d_{6}) & \frac{1}{2}(d_{5}\!-\!3d_{6}) \\ 
  0 & 0 & 0 & 0 & 0 \\ 
  0 & 0 & 0 & 0 & 0 \\ 
  \frac{3}{2}(d_{5}\!+\!d_{6}) & \frac{1}{2}(d_{5}\!-\!3d_{6}) & \frac{1}{2}(d_{5}\!+\!3d_{6}) & 
  -\frac{3}{2}(d_{5}\!-\!d_{6}) & \frac{1}{2}(d_{5}\!-\!3d_{6})  
\end{array}
\right) \;, \\[\MySep] 
C_{d,57}^{80} \!&=&\! (C_{d,57}^{08})^{T}\,,\\[\MySep]
C_{d,57}^{00} \!&=&\! (4d_{5}+6d_{7})\,\mathcal{M}_{\eta}\,.
\end{eqnarray*}
The reader should note that there is no $C_{d,57}^{88}$ matrix since the according vertex rules 
do not give rise to octet-to-octet transitions. In the above, the entries indicated by the dots 
can be read off from the other entries owing to the symmetry of the respective matrices.

Finally, we specify the Born-term matrices. To shorten the length of the coefficients we denote 
${\cal U}[a,b] := \DxD + a\DxF + b\FxF$.

\begin{eqnarray*}
C_{s}^{88} \!\!\! &=& \!\! \frac{1}{4} \left( 
\begin{array}{ccccc} 
  3{\cal U}[2,1] & {\cal U}[-2,-3]           & {\cal U}[4,3]             & -3{\cal U}[0,-1] & {\cal U}[-2,-3]           \\[\MySep]
  \ldots         & \frac{1}{3}{\cal U}[-6,9] & \frac{1}{3}{\cal U}[0,-9] & -{\cal U}[-4,3]  & \frac{1}{3}{\cal U}[-6,9] \\[\MySep]
  \ldots         & \ldots                    & \frac{1}{3}{\cal U}[6,9]  & -{\cal U}[2,-3]  & \frac{1}{3}{\cal U}[0,-9] \\[\MySep]
  \ldots         & \ldots                    & \ldots                    & 3{\cal U}[-2,1]  & -{\cal U}[-4,3]           \\[\MySep]
  \ldots         & \ldots                    & \ldots                    & \ldots           & \frac{1}{3}{\cal U}[-6,9]
\end{array} 
\right) \;, \\[\MySep]
C_{s}^{08} \!\!\! &=& \!\! -\frac{(2D+3D_{s})}{2\sqrt{2}} \left( \!
\begin{array}{ccccc} 
  0 & 0 & 0 & 0 & 0 \\ 
  D\!+\!F & \frac{1}{3}(D\!-\!3F) & \frac{1}{3}(D\!+\!3F) & F\!-\!D & \frac{1}{3}(D\!-\!3F) \\ 
  0 & 0 & 0 & 0 & 0 \\ 
  0 & 0 & 0 & 0 & 0 \\ 
  D\!+\!F & \frac{1}{3}(D\!-\!3F) & \frac{1}{3}(D\!+\!3F) & F\!-\!D & \frac{1}{3}(D\!-\!3F) 
\end{array} 
\!\right)\,, \\[\MySep]
C_{s}^{80} \!\!\! &=& \!\! (C_{s}^{08})^{T} \; , \\[\MySep]
C_{s}^{00} \!\!\! &=& \!\! \frac{1}{6}(2D+3D_{s})^2 \mathcal{M}_{\eta} \; ,
\end{eqnarray*}

The $C_{u}$ matrix has a more complex structure, with each matrix element constructed as
\begin{equation}
(C_{u})_{jb,ia}(\sqrt{s}) = \sum_{c\,\in\lbrace B\rbrace}(\tilde{C}_{u})_{jb,c,ia} \: \mathcal{B}_{u}^{jb,c,ia}(\sqrt{s})\,,\label{eq:Cuofs}
\end{equation}
where $\mathcal{B}_{u}^{jb,c,ia}(\sqrt{s})$ is a function given explicitly in \ref{app:Cuapprox}.
The r.h.s.~of the previous equation contains a sum over the intermediate baryons labeled by $c$, 
but no summation over the channel (double-)indices $(jb)$, $(ia)$ is implied. The energy dependence 
of the $\mathcal{B}_{u}$ functions is not shown explicitly in the matrix specifications that follow, 
which collect the coefficients $\mathcal{B}_{u}^{jb,c,ia}$ and the couplings $(\tilde{C}_{u})_{jb,c,ia}$
in the channel matrix form $(\ldots)_{jb,ia}$. 

\begin{equation*}
\tilde{C}_{u}^{88} = \frac{1}{4}\left( \!
\begin{array}{ccccc} 
  -{\cal U}[2,1] & {\cal U}[-2,-3]           & -2{\cal U}[-1,0]          & \frac{2}{3}{\cal U}[-3,6] & {\cal U}[-2,-3]           \\[\MySep]
  \ldots         & \frac{1}{3}{\cal U}[-6,9] & \frac{2}{3}{\cal U}[3,0]  & 2{\cal U}[-1,0]           & \frac{1}{3}{\cal U}[-6,9] \\[\MySep]
  \ldots         & \ldots                    & \frac{1}{3}{\cal U}[-6,9] & -{\cal U}[-2,-3]          & \frac{2}{3}{\cal U}[3,0]  \\[\MySep]
  \ldots         & \ldots                    & \ldots                    & -{\cal U}[2,1]            & 2{\cal U}[-1,0]           \\[\MySep]
  \ldots         & \ldots                    & \ldots                    & \ldots                    & \frac{1}{3}{\cal U}[-6,9]
\end{array}
\!\right) \; ,
\end{equation*}
\begin{equation*}
\mathcal{B}_{u}^{88} = \left(
\begin{array}{ccccc} 
  \mathcal{B}_{u}^{\pi N,N,\pi N} & \mathcal{B}_{u}^{\pi N,N,\eta N} & \mathcal{B}_{u}^{\pi N,\Sigma,K\Lambda} & 
                                \mathcal{B}_{u}^{\pi N,\Lambda/\Sigma,K\Sigma} & \mathcal{B}_{u}^{\pi N,N,\eta'N} \\[\MySep]
  \ldots & \mathcal{B}_{u}^{\eta N,N,\eta N} & \mathcal{B}_{u}^{\eta N,\Lambda,K\Lambda} 
                                & \mathcal{B}_{u}^{\eta N,\Sigma,K\Sigma} & \mathcal{B}_{u}^{\eta N,N,\eta'N} \\[\MySep]
  \ldots & \ldots & \mathcal{B}_{u}^{K\Lambda,\Xi,K\Lambda} & \mathcal{B}_{u}^{K\Lambda,\Xi,K\Sigma} & 
                                \mathcal{B}_{u}^{K\Lambda,\Lambda,\eta'N}  \\[\MySep]
  \ldots  & \ldots & \ldots & \mathcal{B}_{u}^{K\Sigma,\Xi,K\Sigma} & \mathcal{B}_{u}^{K\Sigma,\Sigma,\eta'N} \\[\MySep]
  \ldots & \ldots & \ldots & \ldots & \mathcal{B}_{u}^{\eta N,N,\eta'N}
\end{array}\right) \;,
\end{equation*}

\begin{eqnarray*}
\tilde{C}_{u}^{08} &=& C_{s}^{08} \;, \\[\MySep]
\mathcal{B}_{u}^{08} &=& \mathrm{diag}(
  \mathcal{B}_{u}^{\eta'N,N,\pi N},
  \mathcal{B}_{u}^{\eta'N,N,\eta N},
  \mathcal{B}_{u}^{\eta'N,\Lambda,K\Lambda},
  \mathcal{B}_{u}^{\eta'N,\Sigma,K\Sigma},
  \mathcal{B}_{u}^{\eta'N,N,\eta'N}
                                    ) \;, \\[\MySep]
\tilde{C}_{u}^{80} &=& (\tilde{C}_{u}^{08})^{T} \;,\quad \mathcal{B}_{u}^{80} = (\mathcal{B}_{u}^{08})^{T} \; , \\[\MySep]
\tilde{C}_{u}^{00} &=& C_{s}^{00} \;,\quad \mathcal{B}_{u}^{00} = \mathcal{B}_{u}^{\eta'N,N,\eta'N} \;.
\end{eqnarray*}
Here we have inserted the mass of the $\eta'$ and $\eta$ mesons for the flavor-singlet and octet mass, respectively, 
neglecting some contributions of higher order in the mixing amplitudes. We also mention that the $C_{u}^{88}$ 
matrix element for the $N\pi\leftrightarrow\Sigma K$ transitions is even more complicated than in the form 
provided above, approximating the exact expression following from Eq.~(\ref{eq:Cuofs}),
\begin{eqnarray*}
(C_{u}^{88})_{\pi N,K\Sigma} &=& 
  \frac{2}{3}D(D+3F)\mathcal{B}_{u}^{\pi N,\Lambda,K\Sigma}+4F(F-D)\mathcal{B}_{u}^{\pi N,\Sigma,K\Sigma} \\
  & \approx & \frac{2}{3}{\cal U}[-3,6] \,\mathcal{B}_{u}^{\pi N,\Lambda/\Sigma,K\Sigma}  
\end{eqnarray*}
with the intermediate baryon mass in $\mathcal{B}_{u}^{\pi N,\Lambda/\Sigma,K\Sigma}$ taken as an average 
of the $\Lambda$ and $\Sigma$ masses.

\subsection{The isospin $I=3/2$ sector}

In the $I=3/2$ sector, the $2\times 2$ coupling matrices read
\begin{equation*}
C_{d,14}^{88} = \left(
\begin{array}{cc} 
  d_{1}+d_{2}+2d_{4} & d_{2}+d_{3}-d_{1} \\ 
  \ldots             & d_{1}+d_{2}+2d_{4} 
\end{array}
\right) \; ,
\end{equation*}

\begin{eqnarray*}
C_{u}^{88} &=& \frac{1}{2} 
\left(\begin{array}{cc} 
  (D+F)^2 & -\frac{1}{3}(D-F)(D+3F) \\  
  \ldots  & (D+F)^2
\end{array}
\right)  \; , \\[\MySep] 
\mathcal{B}_{u}^{88} &=&  
\left(\begin{array}{cc} 
  \mathcal{B}_{u}^{\pi N,N,\pi N} & \mathcal{B}_{u}^{\pi N,\Lambda/\Sigma,K\Sigma} \\  
  \ldots                      & \mathcal{B}_{u}^{K\Sigma,\Xi,K\Sigma}
\end{array}
\right) \;.
\end{eqnarray*}
Any other remaining coefficients not specified here are the same as those provided 
in Appendix~A of \cite{Cieply:2013sya}.

\section{Treatment of the $u$-channel Born terms}
\label{app:Cuapprox}

Calculating the invariant amplitudes stemming from the $u$-channel Born graphs, we find that 
one must project out the s-wave of
\beq
  \hat{g}^{b,ic}\left(\frac{(\sqrt{s}+m_{c}-m_{a}-m_{b})(m_{a}+m_{c})(m_{b}+m_{c})}{u-m_{c}^2} + (\sqrt{s}+m_{c})\right)\hat{g}^{cj,a}\,,
\eeq{eq:ucombs}
and the p-wave of 
\beq
  \hat{g}^{b,ic}\left(\frac{(\sqrt{s}-m_{c}+m_{a}+m_{b})(m_{a}+m_{c})(m_{b}+m_{c})}{u-m_{c}^2} + (\sqrt{s}-m_{c})\right)\hat{g}^{cj,a}\,
\eeq{eq:ucombp}
to obtain the contribution to $f_{0+}$ from the $u$-channel graphs. The p-wave part 
is suppressed by kinematic prefactors, and we shall omit it in the following. 
In Eqs.~(\ref{eq:ucombs}) and (\ref{eq:ucombp}), a summation over the baryon channels $c$ is implied 
and the numbers $\hat{g}^{jb,c}$ specify the axial couplings for $c\rightarrow jb$, 
e.g.~$\hat{g}^{\pi N,N}=-\sqrt{3}(D+F)/(2F_{0})=:\hat{g}^{N,\pi N}$ in the $I=1/2$ sector. 
For fixed $z=\cos\theta$, with $\theta$ denoting the scattering angle in the c.m.~frame, 
the Mandelstam variable $u$ is given by
\begin{eqnarray}
  u(s,z) &=& m_{a}^2+m_{b}^2-s+2\sqrt{q_{ia}^{2}+M_{i}^{2}}\sqrt{q_{jb}^{2}+M_{j}^{2}}-2q_{jb}q_{ia}z\,,\label{eq:uz}\\
  q_{ia}(s) &\equiv& q_{ia} = \frac{\sqrt{(s-(m_{a}+M_{i})^{2})(s-(m_{a}-M_{i})^{2})}}{2\sqrt{s}} \label{eq:qcms}
\end{eqnarray}
for the transition $ia \rightarrow jb$. It is worth noting that, in the physical region 
we have $s>\mathrm{Max}\,((m_{a}+M_{i})^2,(m_{b}+M_{j})^2)$ and for $u(s,z)$ one gets  
the maximum value $u_{\mathrm{max}}=\mathrm{Min}\,((m_{a}-M_{j})^2,(m_{b}-M_{i})^2)$. 
As long as the baryons are stable with respect to a strong decay ($m_{a}<m_{c}+M_{j}\,,\, m_{b}<m_{c}+M_{i}$), 
or $\hat{g}^{b,ic}\hat{g}^{jc,a}=0$, the singularity at $\sqrt{u}=m_{c}$ is not in the physical region. 
Therefore, there should be a region around the meson-baryon thresholds where the partial-wave 
expansions of the $u$-channel Born terms converge. To proceed, we compute
\begin{eqnarray}
I^{jb,c,ia}_{0}(s) &:=& \int_{-1}^{1}dz\,\frac{P_{\ell=0}(z)}{m_{c}^{2}-u(s,z)} \label{eq:I0bjcai}\\ 
 &\hspace*{-30mm} = & \hspace*{-15mm} \frac{1}{2q_{jb}q_{ia}}\log\left(\frac{s+m_{c}^2-m_{a}^2-m_{b}^2-2\sqrt{q_{ia}^{2}+M_{i}^{2}}\sqrt{q_{jb}^{2}+M_{j}^{2}}+2q_{jb}q_{ia}}{s+m_{c}^2-m_{a}^2-m_{b}^2-2\sqrt{q_{ia}^{2}+M_{i}^{2}}\sqrt{q_{jb}^{2}+M_{j}^{2}}-2q_{jb}q_{ia}}\right)\,.\nonumber
\end{eqnarray}
The s-wave projection of Eq.~(\ref{eq:ucombs}) reads 
\beq
-\hat{g}^{b,ic}\left((\sqrt{s}\!+\!m_{c}\!-\!m_{a}\!-\!m_{b})(m_{a}\!+\!m_{c})(m_{b}\!+\!m_{c})I^{jb,c,ia}_{0}(s) 
-2(\sqrt{s}\!+\!m_{c})\right)\hat{g}^{jc,a} \; ,
\eeq{eq:minuszweiCu}
and we write the s-wave amplitude corresponding to the $u$-channel Born graphs as
\begin{equation}\label{eq:fu}
  f_{0+,u}^{I=1/2}(s) = -\frac{\sqrt{E+m}\,C_{u}\sqrt{E+m}}{F_{\Phi}(8\pi\sqrt{s})F_{\Phi}}\,,
\end{equation}
where the matrix $C_{u}$ is found by forming the appropriate isospin combinations with Eq.~(\ref{eq:minuszweiCu}).
Let us consider this expression for the specific case of $m_{a}=m_{b}=m_{c}=m_{N}$ and $M_{i}=M_{j}=M_{\eta}$. 
At the $\eta N$ threshold, we find
\begin{displaymath}
  I^{N\eta,N,N\eta}_{0}(s_{thr}^{N\eta})=\frac{2}{M_{\eta}(2m_{N}-M_{\eta})}\,.
\end{displaymath}
Adding the $s-$channel exchange term and neglecting the small mixing angle, i.e.~the corrections 
of the $\mathcal{O}(\sin\vartheta)$ order, the threshold Born amplitude for the $\eta N$ scattering 
is found as
\begin{equation}
f_{0+,s}^{N\eta}(s_{thr}^{N\eta})+f_{0+,u}^{N\eta}(s_{thr}^{N\eta}) = -\frac{(D-3F)^2M_{\eta}^2}{48\pi F_{\eta}^2\left(1+\frac{M_{\eta}}{m_{N}}\right)}\left(\frac{1}{2m_{N}\!+\!M_{\eta}} + \frac{1}{2m_{N}\!-\!M_{\eta}}\right) \,.
\end{equation}
However, it is problematic to use Eqs.~(\ref{eq:minuszweiCu}) and (\ref{eq:fu}) as a potential kernel 
in a coupled channels equation. Considering e.g.~the $\eta N$ case, one notes a subthreshold cut from
\begin{displaymath}
s_{1,\eta} = \left(m_{N}-\frac{M_{\eta}^2}{m_{N}}\right)^2 \!\! \approx (0.62\,\mathrm{GeV})^2
  \quad\mathrm{to}\quad s_{2,\eta} = m_{N}^2+2M_{\eta}^2 \! \approx \! (1.22\,\mathrm{GeV})^2\,.
\end{displaymath}
It lies partly in the physical region for $\pi N$ scattering, which the $\eta N$ potential 
communicates with in the coupled channels formalism. The loop graphs of $\pi N$ scattering do not 
suffer from such a cut that also spoils coupled channels unitarity. Thus, the appearance 
of the cut in the physical region of the coupled scattering processes has to be considered 
as an artefact of the on-shell approximation, combined with interchanging the order in which 
the partial-waves summation and the loop integrations are performed, see also Sec.~5.2.1 
of \cite{Nissler:2008PhD} for a discussion of this issue. To circumvent this problem, 
we replace the function in Eq.~(\ref{eq:minuszweiCu}) by an approximation which is completely 
free of this near-threshold singularity. This approximation is denoted as $\mathcal{B}_{u}$  
and is constructed explicitly as follows.

\begin{figure}[htb]
\centering
\includegraphics[width=0.48\textwidth]{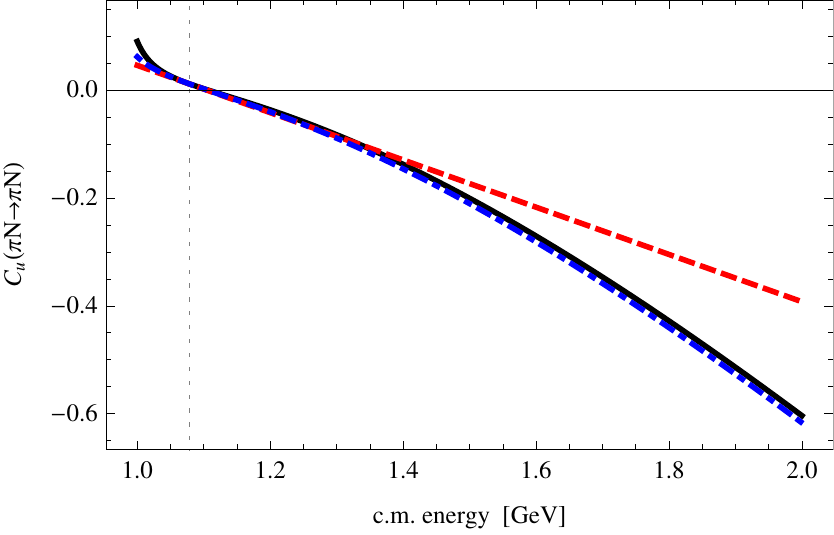}\hspace*{2mm}
\includegraphics[width=0.48\textwidth]{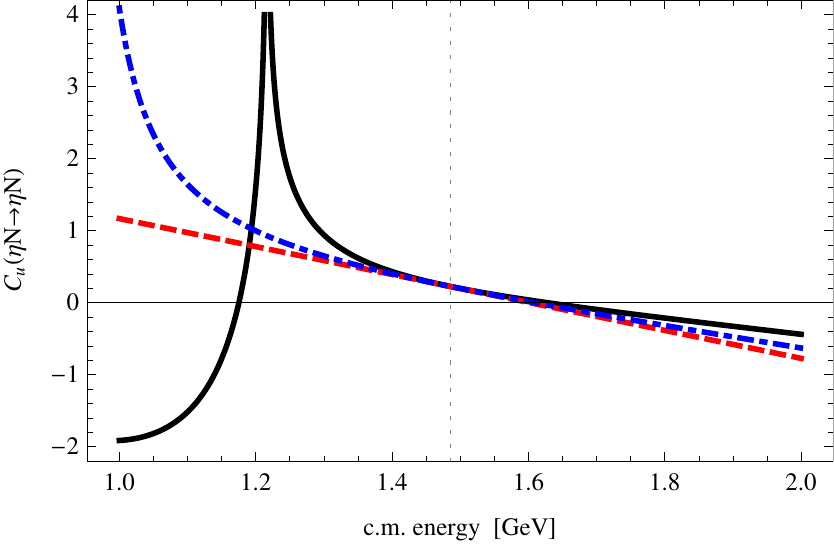} \\
\caption{The energy dependence of the $u$-term Born amplitudes $C_{u}$ for the $\pi N$ (left panel) 
and $\eta N$ (right panel) elastic processes. The continuous black line shows the exact expression, 
the dot-dashed blue line our ${\mathcal B}_{u}$ approximation, and the dashed red line the linear 
approximation (without the $h_{u}$-term). The dotted vertical gray line marks the $\pi N$ 
(or the $\eta N$) threshold.}
\label{fig:CuTilde}
\end{figure}

Suppressing the group-theoretic coupling constant factors $g$, we have
\beq
C_{u}(\sqrt{s}) = \frac{1}{2}(\sqrt{s}\!+\!m_{c}\!-\!m_{a}\!-\!m_{b})(m_{a}\!+\!m_{c})(m_{b}\!+\!m_{c})I^{jb,c,ia}_{0}(s) 
                  -(\sqrt{s}\!+\!m_{c}) \; ,
\eeq{eq:Cuohnegg}
compare Eqs.~(\ref{eq:I0bjcai}) and (\ref{eq:minuszweiCu}). Let us write the approximation as 
\beq
\mathcal{B}_{u}(\sqrt{s}) = C_{u}^{thr} + (\sqrt{s}-\sqrt{s_{thr}})C_{u}'^{\,thr} + (\sqrt{s}-\sqrt{s_{thr}})^2h_{u}(\sqrt{s})\,,
\eeq{eq:Bu}
where
\begin{eqnarray*}
  C_{u}^{thr} \!\!\! &=& \!\!\! C_{u}(m_{b}+M_{j}) \\
              \!\!\! &=& \!\!\! (m_{c}\!-\!m_{a}\!+\!M_{j})(m_{a}\!+\!m_{c})(m_{b}\!+\!m_{c})\frac{1}{Q_{1}} 
                               -(m_{b}\!+\!m_{c})-M_{j} \; ,\\
  C_{u}'^{\,thr} \!\!\! &=& \!\!\! C_{u}'(m_{b}\!+\!M_{j})  \\
                 \!\!\! &=& \!\!\! (m_{a}+m_{c})(m_{b}+m_{c})\frac{1}{Q_{1}} -1 \\
                 \!\!\! & & \!\!\! +\; (m_{c}\!-\!m_{a}\!+\!M_{j})(m_{a}\!+\!m_{c})(m_{b}\!+\!m_{c})
                                   \left(\frac{8m_{b}M_{j}\tilde{q}^2}{3\sqrt{s_{thr}}Q_{1}^3} 
                                   -\frac{2\sqrt{s_{thr}}Q_{2}}{Q_{1}^2}\right) \; ,\\
  Q_{1} \!\!\! &=& \!\!\! \frac{1}{\sqrt{s_{thr}}}\left( M_{j}(m_{b}^2+m_{c}^2+m_{b}M_{j}-M_{i}^2)+m_{b}(m_{c}^2\!-\!m_{a}^2)\right) \; , \\
  Q_{2} \!\!\! &=& \!\!\! \frac{1}{2\sqrt{s_{thr}}^3}\left(m_{b}(m_{a}^2\!+\!m_{b}^2\!-\!M_{i}^2) + M_{j}(3m_{b}\sqrt{s_{thr}}-m_{a}^2+M_{i}^2+M_{j}^2)\right)\,,\\
  \tilde{q}^2 \!\!\! &=& \!\!\! \frac{1}{4s_{thr}}(s_{thr}-(m_{a}+M_{i})^2)(s_{thr}-(m_{a}-M_{i})^2)\,.
\end{eqnarray*}
Here, we shall assume w.l.o.g. that the reaction threshold $\sqrt{s_{thr}}=m_{b}+M_{j}\geq m_{a}+M_{i}$ 
(otherwise, $m_{a}+M_{i}$ would be called the reaction threshold). Now we adjust the function $h_{u}$ 
so that $\mathcal{B}_{u}(\sqrt{s})=C_{u}(\sqrt{s})+\mathcal{O}(p^2)$, where $p$ counts a small chiral 
quantity like a pseudoscalar-meson mass (recall that baryon mass differences are also booked 
as $\mathcal{O}(p^2)$ in the chiral counting). From this requirement, we find
\begin{eqnarray}
h_{u}(\sqrt{s}) \!\!\! &=& \!\!\! 
  \frac{m_{b}^4\!-\!s^2\!-\!\sqrt{s}^3(m_{b}\!+\!m_{c})\!+\!2\sqrt{s}\,m_{b}^3 \!+\! s(m_{a}\!+\!m_{c})(m_{b}\!+\!m_{c})\log\left(\frac{s}{m_{c}^2}
  \right)}{(\sqrt{s}-m_{b})^3(\sqrt{s}+m_{b})^2}\nonumber \\
                \!\!\! &+& \!\!\! 
  M_{j}\frac{5\sqrt{s}^3-21sm_{b}+15\sqrt{s}\,m_{b}^2+m_{b}^3}{3m_{b}(\sqrt{s}-m_{b})^4}\nonumber \\
                \!\!\! &+& \!\!\! 
  M_{j}\frac{2s(m_{a}\!+\!m_{c})(m_{b}\!+\!m_{c}) \log\left(\frac{s}{m_{c}^2}\right)}{(\sqrt{s}\!-\!m_{b})^4 
    (\sqrt{s}\!+\!m_{b})^2} \!+\! \frac{4M_{i}^2}{3M_{j}m_{b}(\sqrt{s}\!-\!m_{b})} \;. 
\end{eqnarray}
The resulting function $\mathcal{B}_{u}(\sqrt{s})$ has a singularity only at $\sqrt{s}=m_{b}$, 
which is below all the reaction thresholds considered here. The approximation is quite reasonable 
as the Fig.~\ref{fig:CuTilde} shows.

Unfortunately, in some cases (e.g.~for $K\Lambda\rightarrow K\Lambda$ with a $\Xi$ in the $u$-channel) 
the approximation deviates strongly from the full result shortly below the reaction threshold, 
well above the singularity of $C_{u}(\sqrt{s})$. For this reason, we decided to use below the channel thresholds 
a more simple approximation employed in Eq.~(17) of \cite{Cieply:2013sya} and match it at the threshold 
to the one given by Eq.~(\ref{eq:Bu}). This roughly corresponds to dropping the $h_{u}$-term below the thresholds.

\section*{References}
\bibliography{mojeCitace}



\end{document}